\documentclass[amsmath,amssymb,prd,twocolumn,nofootinbib]{revtex4}
\usepackage{mathrsfs}
\usepackage{graphicx}
\usepackage{epsfig}
\usepackage{citesort}

\newcommand{\be}{\begin{equation}}
\newcommand{\ee}{\end{equation}}
\newcommand{\bea}{\begin{eqnarray}}
\newcommand{\eea}{\end{eqnarray}}
\newcommand{\bean}{\begin{eqnarray*}}
\newcommand{\eean}{\end{eqnarray*}}
\newcommand\ov[1]{\frac{1}{#1}}
\newcommand\new {\nonumber \\}

\newcommand{\sgrad}{{}^{(3)}\nabla}
\newcommand\fracd[1]{\frac{#1'}{#1}}

\newcommand{\curl}{\,\mbox{curl}\,}

\newcommand{\clp}{{\mathcal{P}}}

\newcommand{\cle}{{\mathcal{E}}}

\newcommand{\clz}{{\mathcal{Z}}}

\newcommand{\clh}{{\mathcal{H}}}
\newcommand{\clv}{{\mathcal{V}}}

\newcommand{\mpc}{\mbox{\rm \,Mpc}}

\newcommand{\nelec}{n_{e}}

\newcommand{\qk}{Q^{(k)}}

\newcommand{\qka}{Q^{(k)}_{a}}

\newcommand{\qkab}{Q^{(k)}_{ab}}

\newcommand{\bmath}{\mathbf} %\bmath may exit in some kind of cls and sty

\begin{document}
%\title{Variables in CAMB}
\title{Cosmic microwave background with Brans-Dicke
           Gravity: I. Covariant Formulation}
\author{Feng-Quan Wu }
\email{wufq@bao.ac.cn}
\author{Li-e Qiang}
\author{Xin Wang}
\author{Xuelei Chen}
\email{xuelei@cosmology.bao.ac.cn}

\affiliation{National Astronomical Observatories, Chinese
Academy of Sciences, \\
20A Datun Road, Chaoyang District, Beijing 100012, China}
\date{\today}
\begin{abstract}
In the covariant cosmological perturbation theory, a 1+3
decomposition ensures that all variables in the frame-independent
 equations are covariant, gauge-invariant and have clear
physical interpretations. We develop this formalism in the case of
Brans-Dicke gravity, and apply this method to the calculation of
cosmic microwave background (CMB) anisotropy and large scale
structures (LSS). We modify the publicly available Boltzmann code
CAMB to calculate numerically the evolution of the background and
adiabatic perturbations, and obtain the temperature and polarization
spectra of the Brans-Dicke theory for both scalar and tensor mode,
the tensor mode result for the Brans-Dicke gravity are obtained
numerically for the first time. We first present our theoretical
formalism in detail, then explicitly describe the techniques used in
modifying the CAMB code. These techniques are also very useful to
other gravity models. Next we compare the CMB and LSS spectra in
Brans-Dicke theory with those in the standard general relativity
theory. At last, we investigate the ISW effect and the CMB lensing
effect in the Brans-Dicke theory.  Constraints on Brans-Dicke model
with current observational data is presented in a companion paper
\cite{wu:2008brans.constraint}(paper II).
\end{abstract}
%\pacs{}
 \maketitle

\section{Introduction}

 The Jordan-Fierz-Brans-Dicke
theory
\cite{Jordan:1949nature,Jordan:1959eg,Fierz:1956,Brans:1961sx,Dicke:1961gz}
(hereafter the Brans-Dicke theory for simplicity) is a natural
alternative and a simple generalization of Einstein's general
relativity theory, it is also the simplest example of a
scalar-tensor theory of gravity
\cite{Bergmann:1968ve,Nordtvedt:1970uv,Wagoner:1970vr,Bekenstein:1977rb,Bekenstein:1978zz}.
In the Brans-Dicke theory, the purely metric coupling of matter with
gravity is preserved, thus ensuring the universality of free fall
(equivalence principle) and the constancy of all non-gravitational
constants. From early on, testing the Brans-Dicke theory with CMB
anisotropy has been considered \cite{Peebles:1970ag}.  However, the
usual approach is to use a metric-based and gauge-dependent method,
i.e. making the calculation with a particular gauge, see. e.g.,
Ref.~\cite{Nariai:1969vh,Baptista:1996rr,chenxuelei:1999brans,Nagata:2002tm,Hwang:1996np}.

The covariant approach to general relativity is an elegant solution to the
``gauge problem", which has plagued the study of linear perturbation in
gauge-dependent methods since the pioneering work of
Lifshitz\cite{Lifshitz:1946}. Before this problem was recognized,
contradictory predictions of the behavior of perturbation of
Friedmann-Lema\^{i}tre-Robertson-Walker (FLRW) cosmologies were made.
In 1980, Bardeen reformulated
the metric approach using gauge-invariant variables \cite{Bardeen:1980kt}
(see also Ref.~\cite{Mukhanov:1990me} for a review on the variables which
has been widely used in recent perturbation calculations).
However, as
pointed out by Ellis \cite{Ellis:1989jt}, although the Bardeen variables are
related to the density perturbations, they are not those perturbations
themselves, since they include metric tensor Fourier components and
other quantities in cunning combinations.  The physical meaning of
Bardeen's gauge-invariant variables are not always transparent.
As emphasized by Hawking \cite{Hawking:1966qi}, the metric tensor
can not be measured directly, so it is not surprising that the variables
used in the metric-based method do not always have a clear physical interpretation.

The covariant approach to general relativity and cosmology has its
origins in the work of Heckmann, Sch\"{u}cking, Raychaudhuri and
Hawking \cite{Heckmann:1955,Raychaudhuri:1957,Hawking:1966qi}.
In 1989, Ellis and Bruni proposed using the spatial gradient of matter
density ($D_a \rho$) as the basic variable to describe density
perturbations \cite{Ellis:1989jt}.  Subsequently, the cosmological
applications have been developed extensively by Ellis and others in
recent
years \cite{Ellis:1990gi,Ellis:2002tq,Ellis:1998ct,Bruni:1992dg,Dunsby:1991xk,Maartens:1996hb,Maartens:1997fg,Maartens:2000fg,Tsagas:1997vf,Tsagas:1999tu,Tsagas:2004kv,Tsagas:2007yx,Li:2006vi,Li:2007jm,Li:2007xw,vanElst:1998rc,Leong:2001qm}.
The method also has been applied to problems in CMB
physics \cite{Challinor:2000as,Challinor:1999xz,Challinor:1999zj,Maartens:1998xg}.
Instead of using the components of metric as basic variables,  the
covariant formalism performs a 1+3 split of the Bianchi and Ricci
identities, using the kinematic quantities, energy-momentum tensors
of the fluid(s) and the gravito-electromagnetic parts of the Weyl
tensor to study how perturbations evolve. The most notable advantage of
this method is that the covariant variables have transparent
physical definitions, which ensures that predictions are always
straightforward to interpret physically. Other advantages include
the unified treatment of scalar, vector and tensor modes, a
systematic linearization procedure which can be extended to consider
higher-order effects (this means the covariant variables are exactly
gauge-invariant,  independent of any perturbative expansion), and
the ability to linearize about a variety of background models, e.g. either the
FLRW or the Bianchi models \cite{Challinor:1998aa,Challinor:1998xk}.

A pioneering work in applying the covariant approach to Brans-Dicke theory is
Ref.~\cite{Hirai:1993cn},
in which a conformal transformation was performed, and calculation
was done in the Einstein frame. More recently,
Ref.~\cite{Carloni:2006gy,Carloni:2006fs} chose the
effective fluid frame,  implying $D_a \phi=0$ and $\omega_{ab}=0$,
i.e. their foliation selects vorticity-free spacelike hypersurfaces
in which $\phi=$ const, hence greatly simplifies the calculations.

The aim of this paper is to construct a full set of covariant and
gauge-invariant linearized equations to calculate angular power spectra of
CMB temperature and polarization anisotropies
in the cold dark matter frame, and to show that the
covariant method  will lead to a clear, mathematically well-defined
description of the evolution of density perturbations. 
In a companion
paper \cite{wu:2008brans.constraint} (heretofore denoted paper II),
we shall apply the formalism developed in this paper to
the latest CMB and large scale structure
data to obtain constraint on the Brans-Dicke parameter.

In \S 2, we briefly review the Brans-Dicke theory and its background
cosmological evolution. The formalism of covariant perturbation theory
is presented in \S 3, and the numerical implementation in \S 4.
We discuss the results on primary anisotropy in \S 5. 
The Integrated Sachs-Wolfe effect and 
gravitational lensing is discussed in \S 6.
Finally, we summarize and conclude in \S 7.

 Throughout this paper we
adopt the metric signature $(+ - - -)$. Our conventions for the
Riemann and Ricci tensor are fixed by $[\nabla_{a},\nabla_{b}] u_{c}
= { R_{abcd}} u^{d}$, where $\nabla_a$ denotes the usual covariant
derivative, and $ R_{ab} \equiv {R_{acb}}^{c}$. We use $\partial_a$
to represent ordinary derivative.  The spatially projected part of
the covariant derivative is denoted by $D_a$. The index notation
$A_l$ denotes the index string $a_1...a_l$.  Round brackets around
indices denote symmetrization on the indices enclosed, square
brackets denote anti-symmetrization, and angled brackets denote the
projected symmetric and tracefree (PSTF) part. We adopt $\kappa=8\pi
G$ and use units with $\hbar=c=k_{B}=1$ throughout. In the
numerical work we use $\mpc$ as unit for distance.

%%!!

\section{The Brans-Dicke Theory and Background Cosmology}

The Brans-Dicke theory is a prototype of the scalar-tensor theory of
gravity. One of its original motivations is to realize
Mach's principle of inertia \cite{Brans:1961sx,Dicke:1961gz}.
It introduced a new degree of
freedom of the gravitational interaction in the form of a scalar field
non-minimally coupled to the geometry.
 The action for the Brans-Dicke theory in the usual (Jordan)
frame is
\begin{equation}
\label{action} {\mathcal S}=\frac{1}{16\pi} \int d^4 x
\sqrt{-g}\left[-\phi R+
\frac{\omega}{\phi}g^{\mu\nu}\nabla_{\mu}\phi \nabla_{\nu}\phi
 \right] +{\mathcal S}^{(m)},
\end{equation}
where $\phi$ is the Brans-Dicke field,  $\omega$ is  a dimensionless
parameter,   and ${\mathcal S}^{(m)}$ is the action for the ordinary
matter fields ${\mathcal S}^{(m)}= \int d^4 x \sqrt{-g} {\mathcal L}^{(m)}.$
Matter is not directly coupled to $\phi$, in the sense that the
Lagrangian density ${\mathcal L^{(m)}}$ does not depend on $\phi$.
For convenience, we also define a dimensionless field
\begin{equation}
\varphi=G \phi,
\end{equation}
where $G$ is the Newtonian gravitational constant measured today.
The Einstein field equations are then generalized to
\begin{eqnarray} \label{generalized Einstein equations}
G_{\mu\nu}&=&\frac{8\pi G}{\varphi}T^{(m)}_{\mu\nu} +
\frac{\omega}{\varphi^2} (\nabla_\mu \varphi \nabla_\nu \varphi
-\frac{1}{2}g_{\mu\nu} \nabla_\lambda \varphi \nabla^\lambda \varphi
)  \nonumber \\ &&
 +\frac{1}{\varphi} (\nabla_\mu \nabla_\nu \varphi - g_{\mu\nu}
\nabla_\lambda \nabla^\lambda  \varphi),
\end{eqnarray}
where $T^{(m)}_{\mu\nu}$ is the stress tensor for all other matter except
for the Brans-Dicke field,  and it satisfies  the energy-momentum
conservation equation, $\nabla^\mu T^{(m)}_{\mu\nu}=0.$
 The equation of motion for $\varphi$ is
\begin{equation}
\nabla_a\nabla^a \varphi= \frac{\kappa}{2\omega+3}  T^{(m)} ,
\end{equation}
here $T^{(m)}=T^{(m) \mu}_{\mu}$ is the trace of the energy-momentum
tensor. The action (\ref{action}) and  the field equation (\ref{generalized
Einstein equations}) suggests that the Brans-Dicke field $\phi$
plays the role of the inverse of the gravitational coupling,
$G_{eff}(\varphi)=\frac{1}{\phi}=\frac{G}{\varphi}$,
 which becomes a function of the spacetime point.

For background cosmology, we treat the ordinary matter as the
perfect fluid with the energy density $\rho$ and pressure $P$,
 \be T^{(m)}_{\mu \nu} = (\rho+P) u_{\mu}u_{\nu}  -P g_{\mu \nu}.
 \ee
The equations describing the background evolution are
\begin{equation}
\label{energy conservation} \rho' + 3\clh(\rho+P)=0,
\end{equation}
\begin{equation}
\label{modified FLRW}\clh^2 = \frac{\kappa S^2}{3\varphi} \rho
+\frac{\omega}{6}\left(\fracd{\varphi}\right)^2
-\clh\fracd{\varphi},
\end{equation}
\begin{equation}
\varphi''+2\clh\varphi'=\frac{\kappa S^2 }{2\omega+3} (\rho-3P),
\label{phi eq backgound}
\end{equation}
where the prime denotes derivative with respect to conformal time
$\eta$, S is the scale factor, and $\clh={S'/S}$. General relativity
is recovered in the limits
\begin{equation}
\omega \to \infty, \quad  \varphi' \to 0,  \quad \varphi'' \to 0.
\end{equation}

To recover the value Newton's gravitational constant
today which is determined by Cavendish type experiments,
we also require that the present day value of $\varphi$ is given by
 \bea \varphi_0=\frac{2\omega +4}{2\omega+3}. \label{phi 0}\eea

\section{Perturbation Theory}
\label{section:perturbation theory}
\subsection{The 1+3 covariant decomposition}
\label{subsection:1+3covarianat}

The main idea of the $1+3$ decomposition is to make space-time splits of
physical quantities with respect to the 4-velocity $u^a$ of an
observer. There are many possible choices of the frame, for
example, the CMB frame in which the dipole of CMB anisotropy
vanishes, or the local rest-frame of the matter. These
frames are generally assumed to coincide when averaged on
sufficiently large scale. Here it will be convenient to choose $u^a$
to coincide with the velocity of the CDM component, since $u^a$ is
then geodesic, and acceleration vanishes. From the 4-velocity $u^a$,
we could construct a projection tensor $h_{ab}$ into the space
perpendicular to $u^a$ (the instantaneous rest space of observers
whose 4-velocity is $u^a$):
\begin{equation}
h_{ab} \equiv g_{ab} - u_{a}u_{b},
\end{equation}
where $g_{ab}$ is the metric of the spacetime. Since $h_{ab}$ is a
projection tensor,  it can be used to obtain covariant tensor
perpendicular to $u^a$, and it satisfies
\begin{equation}
h_{ab} = h_{(ab)}, ~~ u^{a} h_{ab} = 0, ~~ h_{a}^{c}h_{cb} = h_{ab},
~~ h^{a}_{a} = 3.
\end{equation}

With the timelike 4-velocity  $u_a$ and its tensor counterpart $h_{ab}$,
one can decompose a spacetime quantity into irreducible timelike and
spacelike parts. For example, we can use $u_a$ to define the
covariant time derivatives of a tensor ${T^{b \ldots c}}_{d\ldots
e}$:
 \be {\dot{T}^{b \ldots c}}_{d\ldots e} \equiv u^a \nabla_a T^{b \ldots c}_{d\ldots
 e},
 \ee
furthermore, we can exploit the projection tensor $h_{ab}$ to define
a spatial covariant derivative $D^{a}$ which returns a tensor which
is orthogonal to $u^{a}$ on every index:
\begin{equation}
D^{a} {T^{b \ldots c}}_{d\ldots e} \equiv h_{p}^{a} h_{r}^{b} \ldots
h_{s}^{c} h_{d}^{t}\ldots h_{e}^{u} \nabla^{p} {T^{r\ldots
s}}_{t\ldots u},
\end{equation}
 If the
velocity field $u^{a}$ has vanishing vorticity,  $D^{a}$ reduces to
the covariant derivative in the hypersurfaces orthogonal to $u^{a}$.
The projected symmetric tracefree (PSTF) parts of vectors and rank-2
tensors are
\begin{equation}
 {V}_{\langle a\rangle}=
h_a{}^bV_b,
 \ee
 \be
 T_{\langle ab \rangle}= h_{\langle
a}{}^{c}h_{b\rangle}{}^d T_{cd}= h_{(a}{}^{c}h_{b)}{}^d T_{cd}-
{1\over3}\,h^{cd} T_{cd}h_{ab}. \label{angled}
\end{equation}
One can also define a volume element for the observer's
instantaneous rest space:
 \be \varepsilon_{abc}= \eta_{abcd} u^d=\varepsilon_{[abc]} \ ,
 \ee
where $\eta_{abcd}$ is the 4-dimensional volume element
($\eta_{abcd}= \eta_{[abcd]}$, $\eta_{0123}=-\sqrt{|g|}$). Note that
$D_c h_{ab}=0=D_a \varepsilon_{bcd}$.   The skew part of a projected
rank-2 tensor is spatially dual to the projected vector
$T_a=\frac{1}{2}\varepsilon_{abc}
 T^{[bc]}$, and any projected second-rank tensor has the
irreducible covariant decomposition
 \begin{equation}
T_{ab}= {1\over3}\,T h_{ab}+ \varepsilon_{abc}T^c+ T_{\langle
ab\rangle}\,,
\end{equation}
 where $T=T_{cd}h^{cd}$ is the spatial trace. In the 1+3 covariant
 formalism, all quantities are either scalars, projected vectors or
 PSTF tensors.
 The covariant decomposition of  velocity gradient are
\begin{equation}
 \label{4-velocity decompostion}
 \nabla_{a}u_{b} = D_a u_b + u_{a} A_{b},\ee
 \be D_a u_b = \omega_{ab} + \sigma_{ab}
+ \frac{1}{3} \theta h_{ab},
\end{equation}
where $\sigma_{ab}={\rm D}_{\langle a}u_{b\rangle}$ is the shear
tensor which satisfies $\sigma_{ab}=\sigma_{(ab)}$, $\sigma_{a}^{a}
= 0$ and $u^{a} \sigma_{ab} = 0$; $\omega_{ab}={\rm D}_{[a}u_{b]}$
is the vorticity tensor, which satisfies $\omega_{ab}=\omega_{[ab]}$
and $u^{a} \omega_{ab} = 0$. One can also define the vorticity
vector $\omega_a=\varepsilon_{abc}\omega^{bc}/2$ (with
$\omega_{ab}=\varepsilon_{abc}\omega^c$). The scalar $\theta \equiv
\nabla^{a} u_{a}=D^a u_a=3H$ is the volume expansion rate, $H$ is
the local Hubble parameter; and $A_{a} \equiv u^{b}\nabla_{b}
u_{a}=\dot{u}_a$ is the acceleration, which satisfies
$u^{a}A_{a}=0$.   We note that the tensor $D_a u_b$ describes the
relative motion of neighbouring observers.  The volume scalar
$\theta$ determines the average separation between two neighbouring
observers.  The effect of the vorticity is to change the orientation
of a given fluid element without modifying its volume or shape,
therefore it describes the rotation of matter flow.  Finally, the
shear describes the distortion of matter flow, it changes the shape
while leaving the volume unaffected \cite{Tsagas:2007yx}.

Gauge-invariant quantities can be  constructed from scalar variables
by taking their projected gradients.  The comoving fractional
projected gradient of the density field $\rho^{(i)}$ of a species
$i$  is the key quantity of covariant method \cite{Ellis:1989jt},
\begin{equation}
X_{a}^{(i)} \equiv \frac{S}{\rho^{(i)}} D_{a} \rho^{(i)},
\end{equation}
which describes the density variation between two neighbouring
fundamental observers. The comoving spatial gradient of the
expansion rate orthogonal to the fluid flow is
\begin{equation}
Z_{a} \equiv S D_{a} \theta,
\end{equation}
which describes perturbations in the expansion. These quantities are
in principle observable, characterizing inhomogeneity in a covariant
way, and vanishes in the FLRW limit.

The matter stress-energy tensor $T^{(m)}_{ab}$ can be decomposed
irreducibly with respect to $u^{a}$ as follows:
\begin{equation}
T^{(m)}_{ab} \equiv \rho u_{a} u_{b} + 2 u_{(a}q_{b)} - P h_{ab} +
\pi_{ab},
\end{equation}
where $\rho\equiv T^{(m)}_{ab} u^{a} u^{b}$ is the density of matter
measured by an observer moving  with 4-velocity $u^a$, $q_{a} \equiv
h_{a}^{b} T^{(m)}_{bc} u^{c}$ is the relativistic momentum density
or heat (i.e. energy) flux and is orthogonal to
$u^{a}$, $P\equiv - h^{ab} T^{(m)}_{ab} /3$ is the isotropic
pressure, and the projected symmetric traceless tensor $\pi_{ab}
\equiv T^{(m)}_{<ab>}$ is the anisotropic stress, which is also
orthogonal to $u^{a}$.
 The quantities $\rho$, $P$, $q_a$, $\pi_{ab}$
are referred to as {\it dynamical quantities} and $\sigma_{ab}$,
$\omega_{ab}$, $\theta$, $A_a$ as {\it kinematical quantities}. In the FLRW
limit, isotropy restricts $T^{(m)}_{ab}$ to the perfect-fluid form, so
the heat flux  $q_a$ and anisotropic stress $\pi_{ab}$ must
vanish.

The remaining first-order gauge-invariant variables which we need
are derived from the Weyl tensor $C_{abcd}$, which is associated to the
long-range gravitational field and vanishes in an exact
FLRW universe due to isotropy. In analogy to the electromagnetic
field, the Weyl tensor can be split into electric and magnetic
parts, denoted by $E_{ab}$ and $B_{ab}$ respectively. They are both
symmetric traceless tensors and orthogonal to $u^{a}$,
\begin{eqnarray}
E_{ab} &\equiv&  C_{acbd} u^{c}u^{d} =E_{<ab>} , \\
B_{ab} &\equiv& -{}^*C_{acbd} u^{c}u^{d} =- \frac{1}{2}
{\varepsilon_{a}}^{ef} C_{efbd} u^{d}
 \nonumber \\
&=&B_{<ab>}.
\end{eqnarray}
Here $*$ denotes the dual, ${}^*C_{acbd}=\frac{1}{2} \eta_{ac}^{\ \
ef} C_{efbd}$.

For the radiation field, we can make a 1+3 covariant decomposition
of the photon 4-momentum as
\begin{equation}
p^a = E(u^a + e^a), \label{eq:photon momentum}
\end{equation}
where $E=p^a u_a$ is the energy of the photon. $e^a$ describes the
propagation direction of photon in the instantaneous rest space of
the observer. The observer can introduce a pair of orthogonal
polarization vectors $(e_1)^a$ and $(e_2)^a$, which are
perpendicular  to $u^a$ and $e^a$,  to form a right-handed
orthonormal tetrad  $\{ u^a, (e_1)^a, (e_2)^a, e^a \}$ at the
observation point.  The (screen) projection tensor is defined as
\begin{equation}
\clh_{ab} = g_{ab}-u_a u_b + e_a e_b,
\end{equation}
which is perpendicular to both $u^a$ and $e^a$,  and satisfies
$\clh_b^a (e_1)^b = (e_1)^a$.

Using the polarization basis vectors, the observer can decompose an
arbitrary radiation field into Stokes parameters  $I(E,e^a)$,
$Q(E,e^a)$, $U(E, e^a)$ and $V(E,e^a)$\cite{Challinor:2000as}.
Therefore one can introduce a second-rank transverse polarization
tensor $P_{ab}(E,e^c)$
\begin{equation}
P_{ab}(e_i)^a (e_j)^b = \frac{1}{2} \left( \begin{array}{cc}
I + Q & U + V \\
U - V & I- Q \end{array} \right), \label{eq:4}
\end{equation}
for $i$ and $j=1,2$, and we have omitted the arguments $E$ and
$e^a$. $P_{ab} \propto E^3 \clh_a^c \clh_b^d f_{cd}$, where $f_{cd}$
is photon distribution function.   Decomposing $P_{ab}(E,e^d)$ into
its irreducible components, one obtains
\begin{eqnarray}
P_{ab}(E,e^d) &=& -\frac{1}{2}I(E,e^d) \clh_{ab} + \clp_{ab}(E,e^d)
 \nonumber \\ &&
 +\frac{1}{2}V(E,e^d) \epsilon_{abc}e^c, \label{eq:5}
\end{eqnarray}
where  the linear polarization tensor $\clp_{ab}(E,e^d)$ satisfies
\begin{equation}
\clp_{ab}(e_i)^a (e_j)^b = \frac{1}{2} \left( \begin{array}{cc}
Q & U \\
U & - Q \end{array} \right).
\end{equation}

 It is convenient to define energy-integrated multipole for total
intensity brightness  and the electric part of the linear
polarization:
\begin{equation}
I_{A_l} = \int_0^\infty \mbox{d} E\,\int \mbox{d}\Omega ~~~ I(E,
e^c) e_{<A_l>} \ ,
\end{equation}
\begin{eqnarray}
 \cle_{A_l} & = & M_l{}^2 \int_0^\infty \mbox{d} E\,  \int \mbox{d}\Omega
 ~~
e_{\langle A_{l-2}} \clp_{a_{l-1} a_l \rangle} (E,e^c) \ ,
\end{eqnarray}
where $e_{A_l}=e_a e_b e_c ... e_{l}$ and $M_l \equiv
\sqrt{2l(l-1)/[(l+1)(l+2)]}$.

\subsection{The linearized perturbation equations}
\label{subsection:linearized equ}

In the 1+3 covariant approach, the fundamental quantities are not
the metric, which is gauge-dependent,
but the kinematic quantities of the fluid, namely the shear $\sigma_{ab}$,
the vorticity $\omega_{ab}$, the volume expansion rate $\theta$ and
the acceleration $A_a$, the energy-momentum of matter and the
gravito-electromagnetic parts of the Weyl tensor.  The fundamental
equations governing these quantities are the Bianchi identities and
the Ricci identities. The Riemann tensor in these equations is
expressed in terms of $E_{ab}$, $B_{ab}$ and the Ricci tensor
$R_{ab}$. The modified Einstein equation connects the
Ricci tensor to the matter energy-momentum tensor.  In the
following, we have linearized all the perturbation
equations.   We should also note that the definitions of covariant
variables do not assume any linearization, and exact equations can
 be found for their evolution.

The first set of equations are derived from the Ricci identities for
the vector field $u^a$, i.e.
\begin{equation}
2\nabla_{[a}\nabla_{b]}u_c= R_{abcd}u^d\,.  \label{Ris}
\end{equation}
Substituting the 4-velocity gradient (\ref{4-velocity decompostion})
and the decomposition of the Riemann tensor,
 and separating out the
time-like  projected part into the trace, the symmetric trace-free and the skew
symmetric parts, we obtain three propagation equations.  The first
propagation equation is the Raychaudhuri equation,
 \begin{eqnarray} &&
\dot{\theta}+\frac{1}{3}\theta ^{2}-D^{a}\dot{u}_{a}+\frac{%
\kappa }{{2\varphi }}(\rho +3P) +
 \nonumber \\ &&
 \frac{1}{2}\Big(2\omega \frac{{\dot{\varphi}^{2}}%
}{{\varphi ^{2}}}+\frac{1}{\varphi } D _{a} D^{a}\varphi
 + \theta \frac{{\dot{\varphi}}}{ \varphi
}+3\frac{{\ddot{\varphi}}}{\varphi } \Big)=0\, ,
 \end{eqnarray}
which  is the key equation of gravitational collapse, accounting for
the time evolution of $\theta$. The second is the vorticity propagation
equation,
 \bea
\dot \omega _{ab}  - D_{[a} \dot u_{b]}  + \frac{2}{3}\theta \omega
_{ab}  = 0 \,.
 \eea
The last one is the shear propagation equation,
 \bea  && \dot \sigma _{<ab>}  +
\frac{2}{3}\theta \sigma _{ab}  - D_{ < a} \dot u_{b > }  + E_{ab} +
\frac{\kappa }{2}\frac{{\pi _{ab} }}{\varphi }
 \nonumber \\ &&
  + \frac{1}{{2\varphi
}}D_{ < b} D_{a > } \varphi
 + \frac{1}{2}\frac{{\dot \varphi
}}{\varphi }\sigma _{ab}  = 0 \, ,
 \label{shear propagation}
 \eea
which describes the evolution of kinematical anisotropies. It shows
that the tidal gravitational field $E_{ab}$ and the anisotropic stress
$\pi_{ab}$ would induce shear directly, and the shear will change
the spatial inhomogeneity of the expansion through the  constraint
equations (\ref{shear constraint}).

The propagation equations are complemented by three constraint
equations, which are spacelike components of Eq.(\ref{Ris}). The first
is the shear constraint,
 \bea &&
D^b \omega _{ab}  + D^b \sigma _{ab}  - \frac{2}{3}D_a \theta  -
\frac{\kappa }{\varphi }q_a  - \omega \frac{{\dot \varphi
}}{{\varphi ^2 }}D_a \varphi
 \nonumber \\ &&
 - \frac{1}{\varphi }(D_a \varphi )^. -
\frac{{\dot \varphi }}{\varphi }\dot u_a  = 0 \, ,
 \label{shear constraint}
 \eea
which shows the relation between the momentum flux $q_a$, the shear
$\sigma_{ab}$ and the spatial inhomogeneity of the expansion. The
second constraint equation is the vorticity divergence identity,
 \bea
D^c (\varepsilon _{abc} \omega ^{ab} ) = 0 \, .
 \eea
The last one is the $B_{ab}$ equation,
 \bea
B_{ab}  + (D^c \omega _{d(a}  + D^c \sigma _{d(a} )\eta _{b)ce} ^{\
\ \ d} u^e  = 0 \,,
 \eea
which shows that
the magnetic Weyl tensor can be constructed from the
vorticity tensor and the shear tensor. With this last equation
$B_{ab}$ may be eliminated from some equations in favor of the
vorticity and the shear.

So far we have only discussed propagation and constraint equations
for the kinematic quantities.  The second set of equations arises
from the Bianchi identities of the Riemann tensor,
\be \nabla_{[e}R_{cd]ab} = 0 \ , \ee
which gives constraint on the curvature tensor and leads to the
Bianchi identities for Weyl tensor after contracting once,
 \be
 \nabla^dC_{abcd}= \nabla_{[b}R_{a]c}+
{1\over6}\,g_{c[b}\nabla_{a]}R\,.  \label{Bianchi}
 \ee
 The $1+3$ splitting of the once contracted Bianchi identities
leads to two propagation and two constraint equations  which are
similar in form to the Maxwell field equations in an expanding
universe,  governing the evolution of the long range gravitational
field. The first propagation equation is the $\dot{E}$-equation,
 \bea &&
\dot E_{ab}  + \theta E_{ab}  + D^c B_{d(a} \eta _{b)ce} ^{\ \ \ d}
u^e  + \frac{\kappa }{{6\varphi }}[3(\rho  + P)\sigma _{ab}  +
  \nonumber \\ &&
3D_{ < a} q_{b > }  - 3\dot \pi _{ab}  - \theta \pi _{ab} ]
  + \frac{1}{2}\sigma _{ab} (\omega  + \frac{3}{2})\frac{{\dot \varphi ^2 }}{{\varphi ^2 }} -
  \nonumber   \\&&
 \frac{1}{6}\frac{{\sigma _{ab} }}{\varphi }D_\mu  D^\mu  \varphi  + \frac{1}{2}(\omega  + \frac{3}{2})\frac{{\dot \varphi }}{{\varphi ^2 }}D_{ < a} D_{b > } \varphi
  \nonumber \\&&
 + \frac{1}{2}\frac{{\dot \varphi }}{\varphi }E_{ab}  + \frac{3}{4}\kappa \frac{{\dot \varphi }}{{\varphi ^2 }}\pi _{ab}  =
 0 \ ,
 \eea
 and the second propagation is the
$\dot{B}$-equation
 \bea &&
\dot B_{ab}  + (\theta  + \frac{{\dot \varphi }}{{2\varphi }})B_{ab}
- \Big[D^c E_{d(a}  +  \frac{\kappa }{{2\varphi }}D^c \pi _{d(a}  +
 \nonumber \\ &&
\frac{1}{{2\varphi }}D^c D_d D_{(a} \varphi
 - \frac{1}{{6\varphi }}D^c D_\mu  D^\mu  \varphi \;h_{d(a} \;\Big]\eta _{b)ce} ^{\ \ \ d} u^e
 \nonumber \\ &&
   =
 0  \ .
 \eea
This pair of equations for electric
and magnetic parts of the Weyl tensor would give rise wavelike behavior for its
propagation: if we take the time
derivative of the $\dot{E}$-equation, commuting the time and spatial
 derivatives of $B$ term  and substituting from the
$\dot{B}$-equation to eliminate $B$, we would obtain a $\ddot{E}$ term
and a double spatial derivatives term, which together give the wave
operator acting on $E$; similarly we can obtain a wave equation for
$B$ by taking time derivative of the $\dot{B}$-equation.
These waves are also subjected to two constraint equations, which emerge
from the spacelike component of the decomposed Eq.(\ref{Bianchi}).
The first constraint is
 \bea
D^b E_{ab}  - \frac{\kappa }{{6\varphi }}( 2D_a \rho  + 2\theta q_a
+ 3D^b \pi _{ab} ) + \frac{{2\kappa }}{3}\rho \frac{{D_a \varphi
}}{{\varphi ^2 }} -
  \nonumber \\
  \frac{\kappa }{2}\frac{{\dot \varphi }}{{\varphi
^2 }}q_a
 - (\frac{\omega }{3} + \frac{1}{2})\frac{{\dot \varphi }}{{\varphi ^2 }}[\frac{4}{3}\theta D_a \varphi  + (D_a \varphi )^.  + \dot u\;\dot \varphi ] =
 0 \ .
 \label{eq:div-E}
  \eea
This is the div E equation, with the source term given by
the spatial gradient of energy density. It can be regarded
as a vector analogue of the Newtonian Poisson equation,
and shows that the scalar modes will
result in a non-zero divergence of $E_{ab}$, and hence a non-zero
gravitational E-field.
The second constraint equation is
 \bea  &&
D^b B_{ab}  - \frac{\kappa }{{2\varphi }}[(\rho  + P)\eta _{ab} ^{\
\ cd} u^b \omega _{cd}  + \eta _{abcd} u^b D^c q^d ] -
  \nonumber\\  &&
\frac{1}{2}\Big[(\omega \frac{{\dot \varphi ^2 }}{{\varphi ^2 }} -
\frac{1}{{3\varphi }}D_\mu  D^\mu  \varphi - \frac{\theta
}{3}\frac{{\dot \varphi }}{\varphi } + \frac{{\ddot \varphi
}}{\varphi })\eta _{ab} ^{\ \ cd} u^b \omega _{cd}  +
  \nonumber \\ &&
\eta _{abcd} u^b \Big(\omega \frac{{\dot \varphi }}{{\varphi ^2
}}D^c D^d \varphi + \frac{1}{\varphi }(D^c D^d \varphi )^.
  \nonumber \\ &&
+ \frac{\theta }{{3\varphi }}D^c D^d \varphi  + \frac{{\dot \varphi
}}{\varphi }D^c \dot u^d \Big) \Big] = 0 \ . \label{div B}
 \eea
This is the div B equation, with the fluid vorticity serving as source
term. It shows that the vector modes will result in non-zero divergence
of $B_{ab}$, and hence a non-zero gravitational B-field.
The above equations are
remarkably similar to the Maxwell equations of the electromagnetism,
so we have chosen to use $E_{ab}$ and $B_{ab}$ as the symbols.

The last set of equations arises from the twice-contracted Bianchi
identities.
Projecting parallel and orthogonal  to $u^a$, we obtain  two propagation
equations,
 \bea \dot \rho  + \theta (\rho  + P) + D_a q^a  = 0  \ ,\eea
 \bea\dot q_a  + \frac{4}{3}\theta q_a  + (\rho  + P)\dot u_a  +
D^b \pi _{ab}  - D_a P = 0 \ ,\eea
 respectively. For perfect fluids, these reduce to
 \bea \dot \rho  + \theta (\rho  + P) = 0 \ , \eea
 \bea   (\rho  + P)\dot u_a  - D_a P  = 0  \ , \eea
which are the energy conservation equation and momentum conservation equation
respectively.

The background field equation for Brans-Dicke field is given in
Eq.(\ref{phi eq backgound}).   The first order covariant and
gauge-invariant perturbation variable of the Brans-Dicke field is
defined as the spatial derivative of the Brans-Dicke field,
 \bea \clv_a \equiv S D_a\varphi  \ .
 \eea
Taking the covariant spatial derivative of Eq.(\ref{phi eq
backgound}), commuting the spatial and time
 derivatives of $\clv$ term, we could obtain the first order
 perturbation equation for Brans-Dicke field after linearization,
 \bea && \clv_a'' +2 \clh \clv_a' +  S Z_a \varphi' + S^2 D_a D^b
 \clv_b
 \nonumber \\ &&
 = \frac{\kappa S^2}{3+2\omega} \sum_i (1 -3 c^{(i) \, 2}_{s})\rho^{(i)}
 X^{(i)}  \ .
 \label{brans-dicke field perturbation}
 \eea
where the upper index $(i)$  labels the particle species.

In the absence of rotation, $\omega_{ab}=0$, one can define a global
3-dimensional spacelike hypersurfaces that are everywhere orthogonal
to $u^a$. This 3-surfaces is meshed by the instantaneous rest space
of comoving observers. The geometry of the hypersurfaces is
determined by the 3-Riemann tensor defined by
 \bea [D_{a},D_{b}] u_{c}
= { {}^{(3)}R_{abdc}} u^{d}  \ ,
 \eea
which is similar to the definition of Riemann tensor $R_{abdc}$ but
with a conventional opposite sign. The relationship between
${}^{(3)}R_{abdc}$ and $R_{abdc}$ is
 \begin{eqnarray}  {}^{(3)}R_{abcd}
&=& -h_a{}^qh_b{}^sh_c{}^fh_d{}^pR_{qsfp}- v_{ac}v_{bd}+
v_{ad}v_{bc} \nonumber  \\
 &=&  {}^{(3)}R_{[ab][cd]}
\,, \label{3Riemann1}
 \end{eqnarray}
where $v_{ab}={\rm D}_bu_a$ is the relative flow tensor between two
neighbouring observers. In analogy to 4-dimension, the projected
Ricci tensor and Ricci scalar are defined by
 \be
{}^{(3)}R_{ab}={}^{(3)}R_{acbd}  h^{cd}= {}^{(3)}R^c{}_{acb} \ ,
 \ee
and
 \bea
 {}^{(3)}R= {}^{(3)}R_{ab} h^{ab} \,. \label{3Ricci}
 \eea
The ${}^{(3)}R_{ab}$ is determined by the Gauss-Codacci formula
 \be
{}^{(3)}R_{ab}= \frac{1}{3} {}^{(3)}R \, h_{ab}  - \frac{1}{3}
\theta \sigma_{ab} -\frac{\kappa}{2}\pi_{ab} + E_{ab} \ ,
 \ee
 where
 \bea {}^{(3)}R = 2(\kappa \rho-\frac{1}{3}\theta^2) \label{3Ricci scalar} \ .
 \eea
The Eq.(\ref{3Ricci scalar}) is also the generalized Friedmann
equation, showing how the matter tensor determines the 3-space
average curvature.

The last first-order  covariant and gauge-invariant variables can be
obtain from the spatial derivative of the projected Ricci scalar,
 \be \eta_a \equiv \frac{1}{2} S D_a {}^{(3)} R \ ,
 \ee
after a tedious calculation, we obtain
 \bea \eta_a &=& \kappa \frac{\rho X_a}{\varphi}- \kappa \frac{\rho
 \clv_a}{\varphi^2} -\ov{S}(2\clh+\frac{\varphi'}{\varphi})Z_a
 \nonumber \\ &&
 +\ov{S^2}(\omega\frac{\varphi'}{\varphi^2}-\frac{3\clh}{\varphi})(\clv_a'-\clh
 \clv_a) \nonumber\\ &&
+ (\omega+3) \ov{S^2}\frac{\varphi'}{\varphi^2}\clh \clv_a +
 \ov{S}(\omega\frac{\varphi'^2}{\varphi^2}-3\clh\frac{\varphi'}{\varphi})W_a
        \nonumber
 \\ &&
  -
 \frac{\omega}{S^2}\frac{\varphi'^2}{\varphi^3}\clv_a
 - \ov{\varphi}D_a D_\nu \clv^\nu
 -\frac{3}{S} \clh^2 \frac{\clv_a}{\varphi} \ .  \label{eta}
 \eea

\subsection{Mode expansion in spherical harmonics}

 In the linear perturbation theory it is convenient to expand the
$O(\epsilon)$ variables in harmonic modes, since it splits the
perturbations into scalar, vector or tensor modes and decouples the
temporal and spatial dependencies of the 1+3 equations.
 This converts the constraint equations into algebraic relations and
 the propagation equations into ordinary differential equations
 along  the flow lines. In this paper we focus on the
 scalar and tensor perturbation modes, since the vector modes would
decay in an expanding universe in the absence of sources such as topological
defects.

\subsubsection{Scalar mode}

For scalar perturbations we expand  in the scalar eigenfunctions
$Q^{(k)}$ of the generalized Helmholtz equation
 \bea S^2 D^a D_a Q^{(k)}=k^2 Q^{(k)} \label{Helmholtz}
 \eea
 at zero order. They are defined so as to be constant along flow lines, i.e.
independent of proper time $\dot{Q}^{(k)}=O(\epsilon)$, and
orthogonal to the fluid 4-velocity $u^a$.

For the $l-$th multipoles of the radiation anisotropy and
polarization, we expand in the rank-$l$ PSTF tensor, $Q^{(k)}_{A_l}$,
derived from the scalar harmonics with
 \bea Q^{(k)}_{A_l}=\left(\frac{S}{k} \right)^l D_{<a_1...}D_{a_l>}
 Q^{(k)} \label{Q_al} \ ,
 \eea
where the index notation $A_l$ denotes the index string $a_1...a_l$.
The recursion relation for the $Q^{(k)}_{A_l}$,
 \bea Q^{(k)}_{A_l}=\frac{S}{k} D_{<a_l} Q^{(k)}_{A_{l-1}>}
 \label{recursion}
 \eea
follows directly from Eq.(\ref{Q_al}). The factor of $(S/k)^l$ in
the definition of the $Q^{(k)}_{A_l}$ ensures that
$\dot{Q}^{(k)}_{A_l}=0$ at zero-order. The $Q^{(k)}_{A_l}$ also
satisfies some other zero-order properties,
 \bea u^{a_i}Q^{(k)}_{a_1..a_i..a_l}=0, \qquad
 h^{a_i a_j}Q^{(k)}_{a_1..a_i..a_j..a_l}=0.
 \eea
We also have the following differential relations
which can be derived from Eqs.(\ref{Helmholtz}) and
(\ref{recursion}):
\begin{eqnarray}
D^{a_{1}} \qk_{a_{1} a_{2} \ldots a_{l}} &=& {\frac{l}{2l-1}}
{\frac{k}{S}} \left[ 1 - (l^{2}-1) {\frac{K}{k^{2}}}
\right] \qk_{a_{2} \ldots a_{l}} \ ,\\
D^{2} \qk_{a_{1} \ldots a_{l}} &=& {\frac{k^{2}}{S^{2}}} \left[1-
l(l+1) {\frac{K}{k^{2}}} \right] \qk_{a_{1} \ldots a_{l}} \ .
\end{eqnarray}

Now we can expand the gauge-invariant variable in the following
dimensionless harmonic coefficients:
\begin{eqnarray}
X_{a}^{(i)} &=& \sum_{k} k X_{k}^{(i)} \qka  \ ,\\
Z_{a} &=& \sum_{k} {\frac{k^{2}}{S}} Z_{k} \qka  \ , \\
q^{(i)}_{a} &=& \rho^{(i)} \sum_{k} q_{k}^{(i)} \qka  \ ,\\
v^{(i)}_{a} &=& \sum_{k} v_{k}^{(i)} \qka \ ,\\
\pi^{(i)}_{ab} &=& \rho^{(i)} \sum_{k} \pi_{k}^{(i)} \qkab \label{eq_qpi} \ , \\
E_{ab} &=& \sum_{k} {\frac{k^{2}}{S^{2}}} \Phi_{k} \qkab  \ ,\\
\sigma_{ab} &=& \sum_{k} {\frac{k}{S}} \sigma_{k} \qkab \ , \\
A_{a} &=& \sum_{k} {\frac{k}{S}} W_{k} \qka \ , \\
 \clv_a &=& \sum_{k} k \clv_k \qka \ , \\
 \eta_a &=& \sum_{k} \frac{k^3}{S^2} \eta_k \qka \ , \\
 I_{A_l} &=& \rho_\gamma \sum_{k} I_k^{(l)} Q_{A_l}^{(k)} \ , \\
 \cle_{A_l} &=& \rho_\gamma \sum_{k} \cle_k^{(l)} Q_{A_l}^{(k)} \ ,
\end{eqnarray}
where the upper index $(i)$  labels the particle species. The scalar
expansion coefficients, such as $X_{a}^{(i)}$, are first-order
gauge-invariant variables, and their spatial gradients are
second-order, for example $D^a  X_{k}^{(i)}= O(2)$.
In the covariant and gauge-invariant approach, we characterize
scalar perturbations by requiring that the vorticity and the
magnetic part of the Weyl tensor be at most second-order. Demanding
$\omega_{ab}=O(2)$ ensures that density gradients are not from
kinematic effects due to vorticity, and setting
 $B_{ab}=O(2)$ ensures that gravitational waves are
excluded to the first order.

To obtain the scalar equations for the scalar expansion
coefficients, one could substitute the harmonic expansions of the
covariant variables into the propagation and constraint equations
given in the section above.  Here we will consider only the adiabatic modes.
For the (i) fluid,
\begin{equation}
D^{a} P^{(i)} = c_s^{(i) \, 2} \, D^{a} \rho^{(i)},
\end{equation}
where $c_s^{(i)}$ is the adiabatic sound speed of the (i) fluid.
For the spatial gradients of the total density $X_k$, we find
\begin{eqnarray}
&& X_k'+\frac{3 \clh}{\rho} \sum_i \rho^{(i)} X^{(i)}_k \left(c_{s }^{(i)\, 2}
- \frac{P}{\rho}\right) + k\Big[(1+\frac{P}{\rho})Z_k
 \nonumber \\ &&
+ \sum_i
 q^{(i)}_k\Big]  - 3\clh (1+\frac{P}{\rho}) W_k=0 \ .
\end{eqnarray}
For the individual fluid of the $(i)$ species, the propagation
equation satisfies
 \bea && {X'}^{(i)}_k + 3\clh(c^{(i)\, 2}_s - \frac{P^{(i)}}{\rho^{(i)}}) X^{(i)}_k
 +{k}[(1+\frac{P^{(i)}}{\rho^{(i)}})Z_k +q_k^{(i)}]
 \nonumber \\ &&
  - 3\clh(1+\frac{P^{(i)}}{\rho^{(i)}})
 W_k =0 \ .
 \eea
 For the heat fluxes, we have
 \bea && q'^{(i)}_k+ \clh(1-3 \frac{P^{(i)}}{\rho^{(i)}})q^{(i)}_k +(1+\frac{P^{(i)}}{\rho^{(i)}})k W_k +
 \nonumber \\  &&
 \frac{2}{3}k(1-\frac{3K}{k^2}) \pi^{(i)}_k    -k c_s^{(i)\, 2}X^{(i)}_k=0 .
 \label{heat flux}
 \eea
The heat flux for each fluid component is often given by
$q^{(i)}_k=(\rho^{(i)} + P^{(i)})v^{(i)}_k$, so we can derive the
propagation equations for $v^{(i)}_k$ from Eq.(\ref{heat flux}).

We also can obtain the time evolution of the  spatial gradient of
the expansion
 \bea && Z'_k + \clh Z_k + \frac{W_k}{k}(3\clh'-3\clh^2-k^2)
 +\frac{1}{k}\frac{\kappa S^2}{2 \varphi}\sum_i(1+
  \nonumber \\ &&
 3c_s^{(i)\, 2})\rho^{(i)}
 X^{(i)}_k
+\frac{1}{2k} \Big\{\clv_k [-4 \omega \frac{\varphi'^2}{\varphi^3}
-3 \frac{\varphi''}{\varphi^2}
-\frac{S^2\kappa}{\varphi^2}(\rho+
 \nonumber\\ &&
3P)+\frac{k^2}{\varphi}] +4\omega
\frac{\varphi'}{\varphi^2}\clv'_k +3\frac{\clv''_k}{\varphi}
 +k Z_k \frac{\varphi'}{\varphi} +
W_k(4\omega \frac{\varphi'^2}{\varphi^2}
 \nonumber \\ &&
+6
\frac{\varphi''}{\varphi}-3\clh\frac{\varphi'}{\varphi}) +3
\frac{\varphi'}{\varphi}W'_k \Big\}=0 \ .
 \eea

Substituting the covariant harmonic expansion into Eq.(\ref{eta}),
and then taking the time derivative of this equation,   we obtain
the evolution of the spatial gradient of the 3-curvature scalar:
 \bea && k^2\eta'_k = -X_k[S^2
\kappa(\rho+3P)\frac{\clh}{\varphi}+S^2\kappa\rho\frac{\varphi'}{\varphi^2}]
+
\new &&
S^2\kappa\rho\frac{X'_k}{\varphi} +
 \clv_k \Big[
S^2\kappa(\rho+3P)\frac{\clh}{\varphi^2}
+\frac{3}{2}S^2\kappa(\rho-P)\frac{\varphi'}{\varphi^3}
 \new &&
+3\clh\frac{\varphi''}{\varphi^2} -6\frac{\varphi'^2}{\varphi^3}\clh
-(2\omega+\frac{3}{2})\frac{\varphi'\varphi''}{\varphi^3}+2\omega\frac{\varphi'^3}{\varphi^4}
+k^2 \frac{\varphi'}{\varphi^2}
 \new &&
  +\clv'_k \Big[ \frac{S^2
\kappa}{2\varphi^2}(3P-\rho)
 +
(\omega+\frac{3}{2})\frac{\varphi''}{\varphi^2}
 -2\omega\frac{\varphi'^2}{\varphi^3} +6 \clh
 \frac{\varphi'}{\varphi^2}
 \new &&
  -\frac{k^2}{\varphi}  \Big] +
 \clv''_k(\omega\frac{\varphi'}{\varphi^2}-\frac{3\clh}{\varphi})
+W_k (2\omega \frac{\varphi'\varphi''}{\varphi^2} - 2 \omega
\frac{\varphi'^3}{\varphi^3}-
 \new &&
3\clh'\frac{\varphi'}{\varphi}-3\clh\frac{\varphi''}{\varphi}
+3\clh\frac{\varphi'^2}{\varphi^2} ) +
W'_k(\omega\frac{\varphi'^2}{\varphi^2}-3\clh\frac{\varphi'}{\varphi})
-
 \new &&
k Z_k (2\clh'+\frac{\varphi''}{\varphi} -
\frac{\varphi'^2}{\varphi^2})
 -k Z'_k (2\clh+\frac{\varphi'}{\varphi}) \ .
 \label{eq:eta'}
 \eea
As mentioned  by Ref.\cite{Lewis:1999bs},  solving the propagation
equation of $\eta_k$ avoids the numerical instability problem
in isocurvature modes when we work in the CDM frame.

From the shear propagation equation (\ref{shear propagation}), the
propagation equation for $\sigma_k$ becomes
 \bea \sigma'_k &+& \clh \sigma_k - kW_k +k \Phi_k +
 \frac{S^2}{k}\frac{\kappa}{2 \varphi} \rho \pi_k  +
 k\frac{\clv_k}{2\varphi} +
 \nonumber \\ &&
 \frac{1}{2}\frac{\varphi'}{\varphi}\sigma_k=0  \ .
 \eea
From the Div E equation (\ref{eq:div-E}), we could obtain the
$\Phi_k$ equation,
 \bea && 2\frac{k^3}{S^3}(1-3\frac{K}{k^2}) \Phi_k
 -\frac{k}{S}\frac{\kappa \rho}{\varphi}[X_k
 +(1-3\frac{K}{k^2})\pi_k] -
 \new &&
 \frac{3 \clh}{S}\frac{\kappa
 \rho}{\varphi}q_k
 +2\frac{k}{S} \kappa \rho
 \frac{\clv_k}{\varphi^2}
  -\frac{3}{2}\ov{S}\frac{\varphi'}{\varphi^2}\kappa\rho q_k -
  \new &&
(\omega+\frac{3}{2})\frac{k}{S^3}\frac{\varphi'}{\varphi^2}(\clv'_k
+3\clh \clv_k
+ W_k \varphi' )=0 \ .
 \eea
The algebraic equation of $\sigma_k$ can be derived from the shear
constraint equation (\ref{shear constraint})
 \begin{eqnarray}
 \frac{3}{2}k[Z_k &-& \sigma_k(1-3\frac{K}{k^2})]+
 \frac{S^2}{k}\frac{\kappa}{\varphi} \rho q_k
 +\omega\frac{\varphi'}{\varphi^2}\clv_k
  \nonumber \\
 &&+\ov{\varphi}(\clv'_k-\clh\clv_k)
  +\frac{\varphi'}{\varphi}W_k=0 \ .
 \end{eqnarray}
From the first-order
 perturbation equation for the Brans-Dicke field (\ref{brans-dicke field
 perturbation}), we could derive the quadratic differential equation of $\clv_k$
 \bea && \clv''_k + 2\clh \clv_k' +k Z_k \varphi' +k^2 \clv_k
 \nonumber \\
  &&=
\frac{\kappa S^2}{3+2\omega}\sum_i(1 - 3c_s^{(i)\,2})\rho^{(i)}
X^{(i)} \ .
 \label{clv equation}
 \eea
The variables $X_{k}$, $q_{k}$ and $\pi_{k}$ (without upper index
$(i)$ ) refer to variables of the total matter, and can be expressed as
\begin{eqnarray}
\rho X_{k} &=& \rho^{(\gamma)} X^{(\gamma)}_{k} + \rho^{(\nu)}
X^{(\nu)}_{k} + \rho^{(b)} X^{(b)}_{k} + \rho^{(c)} X^{(c)}_{k} \ , \\
\rho q_{k} &=& \rho^{(\gamma)} q^{(\gamma)}_{k} + \rho^{(\nu)}
q^{(\nu)}_{k} + (\rho^{(b)}+p^{(b)}) v^{(b)}_{k}
\nonumber \\
&&+ \rho^{(c)} v^{(c)}_{k}  \  , \\
\rho \pi_{k} &=& \rho^{(\gamma)} \pi^{(\gamma)}_{k} + \rho^{(\nu)}
\pi^{(\nu)}_{k}.
\end{eqnarray}

\subsubsection{Tensor mode}

For tensor modes, we expand the first order perturbation variables in the rank-2,
zero-order PSTF tensor eigenfunctions $Q_{ab}^{(k)}$ of the comoving Laplacian,
\bea
S^2 D^c D_c Q_{ab}^{(k)}=k^2 Q_{ab}^{(k)} .
\eea
Similar to the case of scalar modes, this equation holds at the zero-order.
The tensor harmonics
are transverse,  orthogonal to $u^a$, and constant along the integral curves of $u^a$:
\bea
D^a Q_{ab}^{(k)}=0, ~~~~  u^a Q_{ab}^{(k)}=0, ~~~~ \dot{Q}_{ab}^{(k)}=0\ .
\eea
They can also be classified as having electric parity (denoted by $Q_{ab}^{(k)}$)
or magnetic parity (denoted by $\bar{Q}_{ab}^{(k)}$). These two parity harmonics
are related by a curl:
\begin{eqnarray}
\curl Q^{(k)}_{ab} &=& {\frac{k}{S}} \sqrt{1+ {\frac{3K}{k^2}}}
\bar{Q}^{(k)}_{ab}  \ ,
 \\
\curl \bar{Q}^{(k)}_{ab} &=& {\frac{k}{S}} \sqrt{1+ {\frac{3K}{k^2}}}
{Q}^{(k)}_{ab} \ .
\end{eqnarray}

For tensor mode perturbations, the vorticity and
all gauge-invariant vectors vanish at the first
order, i.e.
$\omega_{ab}, X_{a}, Z_a, q_a, A_a, \clv_a, \eta_a$ all equal to
zero \cite{Dunsby1997CQGra..14.1215D,Challinor:1998aa}.
The rest  rank-2 gauge-invariant tensors are constrained to be transverse:
\begin{eqnarray}
&&\sgrad^{a} E_{ab} = 0 , \quad \sgrad^{a} B_{ab}=0,
 \nonumber \\
&&\sgrad^{a} \sigma_{ab} = 0, \quad \sgrad^{a} \pi_{ab} = 0.
\end{eqnarray}
And they can be expanded  in electric and magnetic parity tensor harmonics:
\begin{eqnarray}
E_{ab} & = &\sum_{k} \frac{k^2}{S^2} (E_k Q^{(k)}_{ab} +
\bar{E}_k \bar{Q}^{(k)}_{ab}) \ , \\
B_{ab} & = &\sum_{k} \frac{k^2}{S^2} (B_k Q^{(k)}_{ab} +
\bar{B}_k \bar{Q}^{(k)}_{ab}) \ , \\
\sigma_{ab} & = & \sum_{k} \frac{k}{S} (\sigma_k Q^{(k)}_{ab}+
\bar{\sigma}_k \bar{Q}^{(k)}_{ab}) \ , \\
\pi^{(i)}_{ab} & = & \rho^{(i)} \sum_{k} (\pi_k^{(i)} Q^{(k)}_{ab}
+ \bar{\pi}_k^{(i)} \bar{Q}^{(k)}_{ab})\ ,
\end{eqnarray}

Substituting these into equations in section
\ref{subsection:linearized equ}, we obtain the $E_k$ and
$\sigma_k$ propagation equations:
\bea &&  k^2 E'_k +k^2
E_k(\clh+\frac{1}{2}\frac{\varphi'}{\varphi}) -
k^3(1+3\frac{K}{k^2})\sigma_k +
 \nonumber \\ &&
 \frac{\kappa S^2}{2 \varphi}(\rho+P)
k \sigma_k + \frac{\varphi'^2}{2\varphi^2} (\omega+\frac{3}{2})k
\sigma_k + \frac{\kappa S^2 \rho \pi_k }{\varphi} \clh
+
 \nonumber \\ &&
\frac{3}{2}\frac{\kappa S^2 P \pi_k}{\varphi} \clh +\frac{3}{4}
\frac{\varphi'}{\varphi^2} \kappa S^2 \rho \pi_k -\frac{1}{2\varphi}
\kappa S^2 \rho \pi'_k =0 \ , \eea \bea \sigma'_k=- \clh \sigma_k -k
E_k - \frac{\kappa}{2k}\frac{S^2 \rho \pi_k}{\varphi} -
\frac{1}{2}\frac{\varphi'}{\varphi} \sigma_k  \, . \eea

\section{Numerical Implementation}
We carry out our numerical study by modifying the CAMB code. The
original CAMB code,  written by Antony Lewis and Anthony
Challinor\cite{CAMB}, is a FORTRAN 90 program which calculates CMB
anisotropies in the standard Einstein general relativity, by solving
the Boltzmann-Einstein equations for various components in the Universe.
Most of the equations to be solved are in the
file {\it equations.f90}, which can be modified conveniently.  The background evolution
equation $d\tau/da$ is written in function {\it dtauda}, and it can
be modified for different background. The Boltzmann-Einstein
equation group is listed in the functions {\it fderivs} (scalar mode for
flat Universe), {\it fderivst} (tense mode for flat Universe), {\it
derivs} (scalar mode for non-flat Universe) and {\it derivst} (tense
mode for non-flat Universe). This equation group includes the
propagation equations of scalar factor $S$,  the 3-Ricci scalar
perturbation $\eta$, the cold dark matter perturbation $X_c$, the baryon
perturbations $X_b$ and $v_b$,  photon multipole moments, and
neutrino multipole moments in the covariant approach. The CAMB code
uses the Runge-Kutta method (subroutine dverk in file
subroutines.f90) to solve these equations. To speed up the calculation,
the line-of-sight integration method first developed by Seljak and
Zaldarriaga\cite{Seljak:1996is} is used: the differential
equation for photon temperature perturbation is integrated along the l.o.s.
to obtain $\delta T/T$. The multipoles today is a definite integral
of source term multiplied by the spherical Bessel functions from early
time to today. The source term of scalar perturbation at a given time
for a given wavenumber is encoded in subroutine {\it output}. The
subroutine evolves the perturbation equations and does the
integration in cmbmain.f90 file. The main routine for
running CAMB is wrapped in file camb.f90.
Using these equations, we modify the code for
calculation in the Brans-Dicke theory. The
most important three parts of modifications are: the
background evolution, the Boltzmann-Einstein differential equations,
and the source term in the line-of-sight integration.

For the background evolution, we implement the procedure described
in the appendix of Ref. \cite{chenxuelei:1999brans}. To satisfy the
end point condition Eq.~(\ref{phi 0}), we start from an epoch which
is deemed early enough. We then evolve the model forwards (to avoid
numerical instability we do not evolve backwards) to obtain the
$\varphi$ value today. The procedure is repeated with a Brent
algorithm (see e.g. \cite{Press}) to find the initial value of
$\varphi$ at that epoch. In doing this we set $\varphi'=0$ and
$\clv_k=\clv'_k=0$  at the initial point. The initial condition
$\varphi'=0$ can be justified by Eq.(\ref{phi eq backgound}):  in
the radiation dominated era, the R.H.S. of Eq.(\ref{phi eq
backgound}), $\rho-3P$ is negligible compared with other terms, then
 \bea  \varphi'=c_1+c_2 S^{-2}  . \eea
This mean that any initial velocity quickly dies out in a few Hubble
times and approaches a terminal velocity $c_1$, this velocity is
constrained by nucleosynthesis, so it should be very small.
The initial condition $\clv_k=\clv'_k=0$ is the simplest
choice which matches the requirement of Eq.(\ref{clv equation}).
Initial perturbations in $\varphi$ are damped during the radiation dominated era,
so the choice of the initial condition of $\clv_k$ have little
impact on CMB anisotropy in the adiabatic perturbation case.

To realize the background evolution described above, we write a
separate module. The function of this module is that, for a given
value of $\varphi$ today which is determined by Brans-Dicke parameter
$\omega$,  we first find out the initial value of $\varphi$ at
sufficiently early time which can evolve the given value of $\varphi$ today,
and then we could calculate $\varphi$ and $\varphi'$ at each scale
factor $S$ and store them into arrays for interpolation in
subsequent process.   Therefore, if one want to use $\varphi$ and
$\varphi'$ in the code, just simply use this module first.

To be consistent with modified Friedmann equation (\ref{modified
FLRW}), in the code we define the critical density as
 \bea  \rho_{cr}=  \frac{3\varphi_0}{\kappa } H_0^2 \ ,
 \eea
where $H_0$ is the hubble parameter today and $\varphi_0$ is given
in Eq.(\ref{phi 0}). This definition differs from the conventional one
by an additional factor $\varphi$. The definition of the fractional
density is the same as the traditional one:
 $ \Omega^{(i)}={\rho_{0}^{(i)}}/{\rho_{cr}} \ .
 $
Because $\varphi'$ approximately vanishes today (c.f.
Fig.(\ref{fig:phi prime evolution})), from Eq.(\ref{modified FLRW})
we find $\Omega_{total}\simeq 1$ for the flat geometry.
This definition is convenient in studying the
non-flat universe. We also should note that the
difference with the traditional one is very small, in most case, less than 1\%,
because $\varphi_0=1.001$ when $\omega=50$.

In this work we adopt the cosmological
constant as dark energy,  this is equivalent to set the potential of
the Brans-Dicke field to  a constant. The more general case of
extended quintessence
\cite{Uzan:1999ch,Amendola:1999qq,Chiba:1999wt,Perrotta:1999am,Holden:1999hm,Baccigalupi:2000je,Chen:2000xxa}
will be dealt with in future studies. Below,
we adopt the $\Lambda$CDM model with Einstein gravity which
best fit the WMAP five-year data\cite{Komatsu:2008hk} as our fiducial
model, i.e. 
\ensuremath{\Omega_b h^2 = 0.02265}, \ensuremath{\Omega_c h^2=
0.1143},
 \ensuremath{H_0 = 70.1}~${\rm
km~s^{-1}~Mpc^{-1}}$,
 \ensuremath{n_s = 0.960},
  \ensuremath{\Delta_{\cal R}^2 =
2.457\times 10^{-9}} at $k=0.002~{\rm Mpc^{-1}}$,
\ensuremath{\tau=0.084} and the equation state of dark energy
$w=-1$.
%\ensuremath{\Omega_\Lambda = 0.721},
%\ensuremath{\Omega_b = 0.0462}, \ensuremath{\Omega_c = 0.233},
%\ensuremath{H_0 = 70.1}~${\rm km~s^{-1}~Mpc^{-1}}$,
% \ensuremath{n_s = 0.960},
%  \ensuremath{\Delta_{\cal R}^2 =
%2.457\times 10^{-9}} at $k=0.002~{\rm Mpc^{-1}}$, \ensuremath{z_{\rm
%reion} = 10.8} and the equation state of dark energy $w=-1$.

For the Boltzmann-Einstein differential equations, we modified
the scale factor evolution equation and $\eta_k$ propagation
equation according to the Eqs. (\ref{modified FLRW}) and
(\ref{eq:eta'}) respectively  in functions {\it fderivs} and {\it
fderivst} in equations.f90.  Some other complementary equations,
such as the constraint equations,  have also been modified
correspondingly.

To speed up the calculation, the CAMB code integrates the system of
differential equations by using the line-of-sight integration
method, first developed by Seljak and
Zaldarriaga for the CMBFAST code \cite{Seljak:1996is}.  In this
method,  the multipole moment of photon intensity $I^{(l)}_k$  could
be express as \cite{lewis2000thesis,Challinor:2000as},
\begin{eqnarray}
I^{(l)}_k &=& 4 \int^{\eta_0}\mbox{d} \eta \,\,
 S e^{-\tau}\Bigg\{\Big(\frac{k}{S} \sigma_k + \frac{1}{4} \nelec
\sigma_{\mbox{\scriptsize T}} \kappa_2{}^{-1}
(\frac{3}{4}I_k^{(2)}
 \new &&
+\frac{9}{2}\cle_k^{(2)}) \Big) \times  \left[\frac{1}{3}
\Phi_l^\nu(x) + \frac{1}{k^2 r^2} \frac{\mbox{d}^2}{\mbox{d} x^2}
\Phi_l^\nu(x)\right]
\new &&
 +\sigma_{\mbox{\scriptsize T}} v_k
\frac{1}{k r } \frac{\mbox{d}}{\mbox{d} x}\Phi_l^\nu(x)
\nonumber \\
&&- \left[\frac{1}{3}\frac{k}{S}\clz_k-\frac{1}{4}\nelec
\sigma_{\mbox{\scriptsize T}} I^{(0)}_k \right] \Phi_l^\nu(x)
\Bigg\}, \label{eq:Intensity}
\end{eqnarray}
where $\eta_0$ is the conformal time today, $x=(\eta_0-\eta)/r$,
$r=1/\sqrt{|K|}$, and $\tau$ is the zero-order optical depth back to $x$.
Here,
\bea \Phi_l^\nu(x)&=&\frac{l!}{(l-\nu)!}\frac{j_l(x)}{x^\nu},
 \eea
are the ultra-spherical Bessel functions,
$\kappa_2=(1-3K/k^2)^{{1}/{2}},$ and $\cle_k^{(2)}$ is the
quadrupole of the E-like polarization of the CMB photons. After
integration by parts, one could eliminate the derivatives of
ultra-spherical Bessel functions and write temperature anisotropies
as a time integral over a geometrical term $\Phi_l^\nu(x)$ and a
source term:
 \bea
I^{(l)}_k &=& 4 \int^{\eta_0}\mbox{d} \eta \,\, \Phi_l^0(x) \times {\mathcal S}
 \eea
 where the source term is given by
 \bea
{\mathcal S} &=& \frac{1}{12 {k}^{2} \kappa_2 }\, \Bigg[12 k \sigma_k''
 e^{-\tau}  \kappa_2 + 24\,k \sigma_k' g(\eta) \kappa_2+
 \nonumber \\ &&
 12\,k\sigma_k{g'(\eta)} \kappa_2
+3 g''(\eta) \zeta_k  +6\,g'(\eta)\zeta_k'+3\,g(\eta)\zeta_k''
 \nonumber \\ &&
+12\,k \,{ \kappa_2} g'(\eta) v_k
 + 12\,k\,{ \kappa_2} g(\eta)v'_k
+4\,{k}^{3} \sigma_k e^{-\tau} \kappa_2 +
 \nonumber \\ &&
 {k}^{2} g(\eta)\zeta_k
-4\,{k}^{3}{ e^{-\tau}} \clz_k{ \kappa_2}+3\,{k }^{2}\,g(\eta)
I_k^{(0)} \kappa_2 \Bigg]   \ ,
 \eea
in which
 \bea  \zeta_k&=&\frac{3}{4}I_k^{(2)}+\frac{9}{2}\cle_k^{(2)} \ , \\
        g(\eta) &=& - \tau' e^{-\tau}= n_e \sigma_{\mbox{\scriptsize T}} S
        e^{-\tau} \ ,
  \eea
$g(\eta)$ is the visibility function. Using the first order
derivative perturbation equations described in Section
\ref{section:perturbation theory}, $\sigma_k''$ and $\zeta_k''$ in
the source terms could  be further expanded to the zeroth and first
order derivative terms which are expressed in variables used in
the {\it output} subroutine in the CAMB code.

\section{Results}

\begin{figure}
 \centering{ \epsfig{file=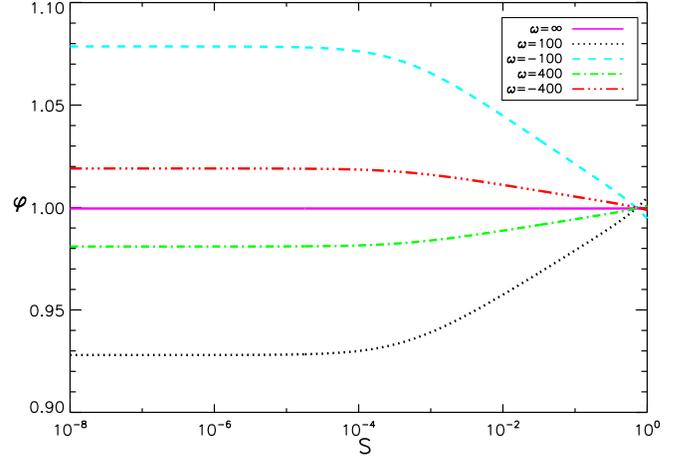,width=3.4in} \caption{The time
evolution of the Brans-Dicke field $\varphi$.} \label{fig:phi
evolution} }
\end{figure}

\begin{figure}
 \centering{ \epsfig{file=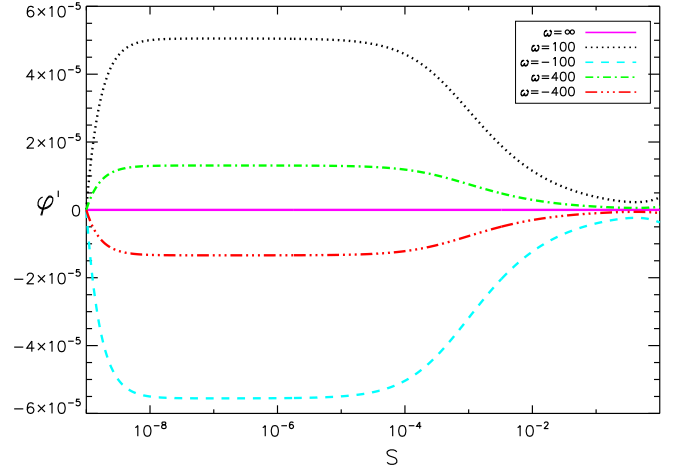,width=3.4in} \caption{The
evolution of time derivative of the Brans-Dicke $\varphi'$.}
\label{fig:phi prime evolution} }
\end{figure}

In Fig.\ref{fig:phi evolution}, we show the time evolution of
the Brans-Dicke field $\varphi$. For models with $\omega>0$, the
value of $\varphi$ increases with time, whereas for models with
$\omega<0$, $\varphi$ decreases with time. During the radiation
dominated era, the variation of $\varphi$ is very small, almost
zero. When entering  the matter dominated epoch, $\varphi$
begins to increase or decrease. After the domination of 
the dark energy, $\varphi$ changes more rapidly.  We also plot the 
time evolution of $\varphi'$ in Fig.\ref{fig:phi prime evolution},
as can be seen from that figure, $|\varphi'|$ reaches a terminal velocity in
the radiation dominated era, and then begin to decay  in the matter
dominated epoch, but as the dark energy becomes dominant, it increases
again, and its present day value for this particular 
model is of the order $10^{-6}$.

\begin{figure}
\centering{ \epsfig{file=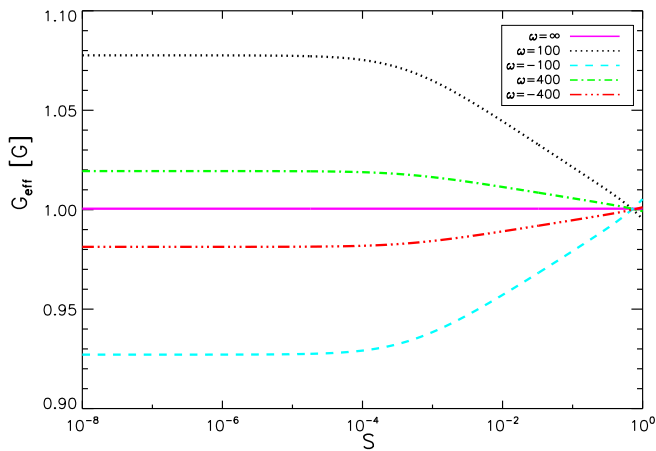,width=3.4in} \caption{The time
evolution of the effective Newtonian gravitational coupling
$G_{eff}$.}
 \label{fig:G evolution} }
\end{figure}

The effective Newtonian
gravitational coupling $G_{eff}$ is the inverse of $\varphi$ in the
unit of $G$.  The time evolutions of $G_{eff}$ are shown in
Fig.\ref{fig:G evolution}. We can see that $G_{eff}$ changes
rapidly at low redshift, so it may not be reliable to use the Type Ia
supernovae (SNe Ia) data to constrain the Brans-Dicke
theory: the Chandrasekhar mass $M_{Ch} \propto G^{-3/2}$, so
the variation of the gravitational coupling $G$ means that the peak
luminosity of SNe, which is approximately proportional to the
Chandrasekhar mass,  may also change, making it not reliable as a standard
candle.

\begin{figure}  \centering{ \epsfig{file=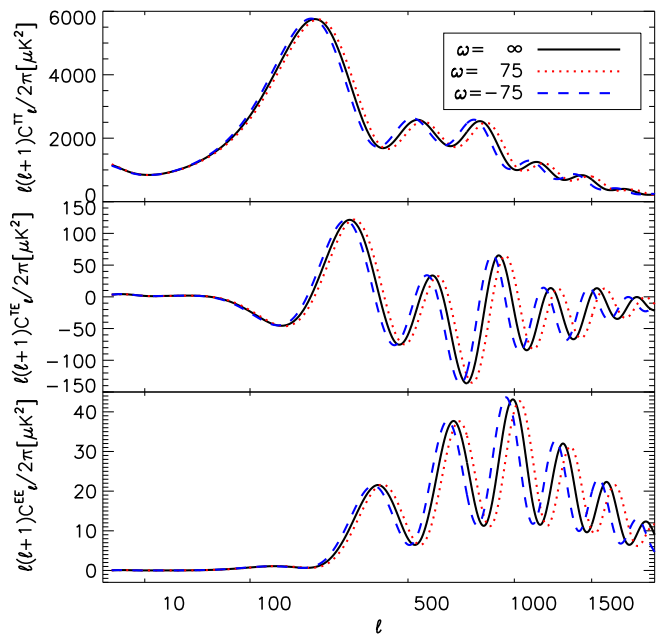,width=3.4in}
\caption{CMB temperature and polarization power spectra for
Brans-Dicke theories with $\omega=\infty, \pm75$ in the scalar mode.}
\label{fig:CMB spectrum} }
\end{figure}

\begin{figure}
\centering{
\epsfig{file=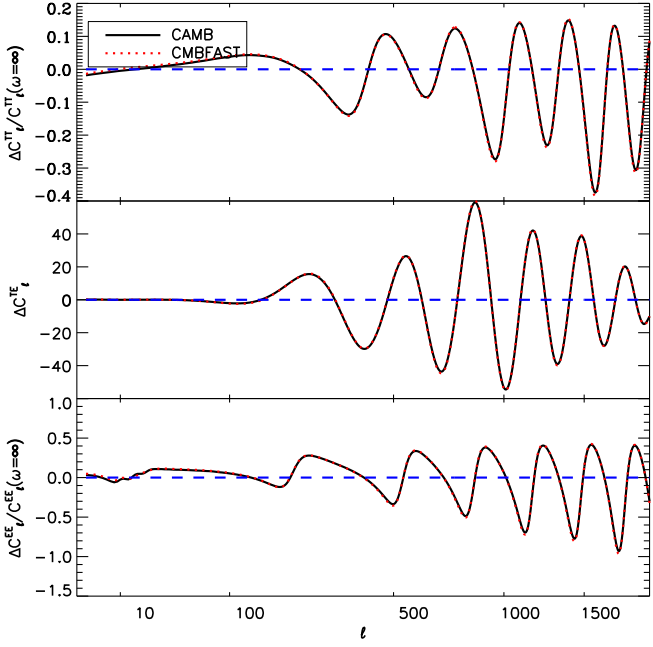,width=3.4in} \caption{$\Delta
C_l=C_l(\omega=\infty)-C_l(\omega=75)$ for TT, TE and EE
correlations.}
 \label{fig:diffs} }
\end{figure}

The CMB angular power spectra for the
Brans-Dicke theories with $\omega=\infty$(i.e. general relativity)
and $\pm75$ are plotted in Fig.\ref{fig:CMB spectrum}, and the resulting
difference are plotted in Fig.\ref{fig:diffs}.  As can be
seen, compared with the general relativity theory with the same
cosmological parameters, both the location and height of the CMB
acoustic peaks are changed. The Brans-Dicke model with a positive
$\omega$ has broader and lower acoustic peaks for this set of
parameters. As $|\omega|$ increases, the difference in CMB angular
spectra between Brans-Dicke theory and general relativity
diminishes. The difference is more apparent at large $l$ (small
angular scale), so high resolution CMB data would be very useful in
distinguishing the different models.  From Fig.\ref{fig:CMB
spectrum}, it is also very clear that the polarization spectra have
a strong discriminating power.  With the higher angular resolution
and polarization data which we expect in the nearby future, we
should be able to lift the the degeneracy of parameters and place a
more stringent constraint on the Brans-Dicke models.

We compare the result of our new code with those
obtained with the CMBFAST code in the synchronous gauge\footnote{There are some
typos in Eqs.(19) and (20) of Ref.~\cite{chenxuelei:1999brans}.
A prime $'$ was missed in the last term of Eq.(19),  i.e. it should
read $\frac{3 a'\chi'}{a\phi}$. A factor of 2 in the denominator was missed in
the last term of Eq.(20), i.e, it should
read $\frac{-1}{2\phi}(\chi'-\frac{a'\chi}{a})$. Most conclusions 
of that paper were not affected, but at small $\ell$ the $C_{\ell}$ was slightly
over-estimated.} in Ref.~\cite{chenxuelei:1999brans}.  
We find that the difference in the CMB
power spectra is typically less than 1
percent and is due primarily to the difference in
the original (Einstein gravity) codes--for really making
highly precise constraint on cosmological parameters with the CMB
data, the precision of the CMB Boltzmann needs to be further improved.
The new code of course has
better program architecture and runs faster. Particularly, if one
calculate $\partial C_l/\partial\omega$, which reflects the impact of the
gravity model on the CMB angular power spectrum, the results of the two code
agree with each other at high precision, as shown in
Fig.\ref{fig:diffs}. The result on $\Delta
C_l=C_l(\omega=\infty)-C_l(\omega=75)$ for TT, TE and EE
correlations are very consistent in two codes, and the two curves
are almost indiscernible in Fig.\ref{fig:diffs}.

\begin{figure}
\epsfig{file=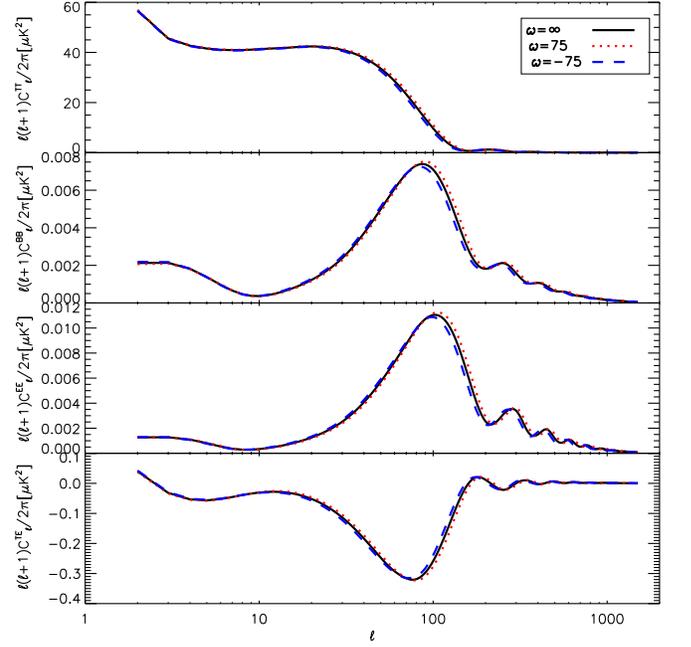,width=3.4in} \caption{ CMB temperature
and polarization power spectra for the Brans-Dicke theories in
tensor mode. The solid, dotted and dashed curves represent the
Brans-Dicke model with $\omega=\infty,75$ and $-75$ respectively.
The tensor-to-scalar ratio R is set to 0.1}
\label{fig:spectrum.tensor}
\end{figure}

We also plot the CMB temperature and polarization spectra yielded by
tensor modes in Fig.\ref{fig:spectrum.tensor}. The tensor-to-scalar
ratio is set to 0.1. The primordial gravitational wave produces
large temperature fluctuations at the large scales, as well as a
unique B mode polarization. In contrast to scalar modes, when
compared with the result of general relativity, the height of the
peaks are higher for positive $\omega$.  Similar to the scalar mode,
positive $\omega$ shifts the peaks to smaller scale.  At both the
very large scales and very small scales, the differences in spectra
between the Brans-Dicke theory and the general relativity are very
small, almost invisible, and the differences are only sensitive at
$l\sim 80$.

\begin{figure}
 \epsfig{file=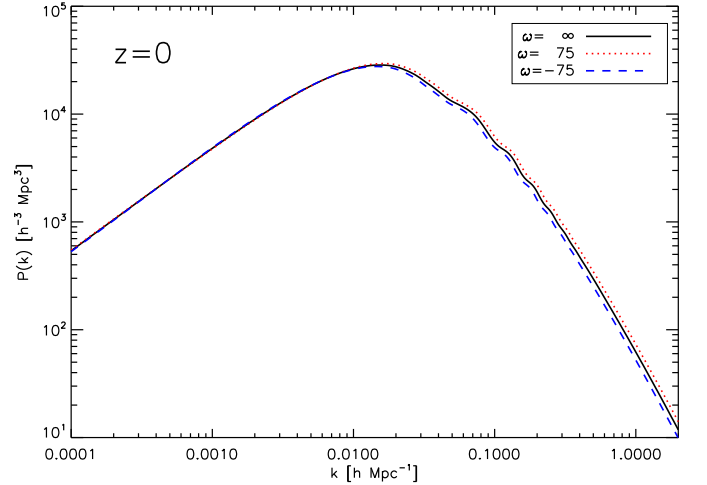,width=3.5in} \caption{ The matter
power spectra at z=0. The solid, dotted and dashed curves represent
the Brans-Dicke model with $\omega=\infty,75$ and $-75$
respectively.} \label{fig:spectrum.matter.z0}
\end{figure}

Fig.\ref{fig:spectrum.matter.z0} shows the impact of Brans-Dicke
field on the matter power spectra at $z=0$. For $\omega=75$, the
bend of the matter power spectrum occurs at short wavelengths, and
there is thus more small-scale power, in agreement with the
prediction of Ref. \cite{Liddle:1998ij}.

\section{Integrated Sachs-Wolfe effect and Gravitational Lensing}

The integrated Sachs-Wolfe (ISW) effect 
is the secondary CMB anisotropy caused by
the time-varying gravitational potential $\Phi$.  CMB temperature
fluctuation of ISW effect in the direction $\mathbf{\hat{n}}$
is given by
\begin{equation}
\delta_{T}^{\rm{ISW}}(\bmath{\hat{n}}) \equiv
\frac{\Delta_T^{\rm{ISW}}(\bmath{\hat{n}})}{T_0} =- 2
\int_{0}^{z_{{LS}}} dz ~ \frac{\partial \Phi}{\partial
z}(\bmath{\hat{n}},z) \label{equa:isw}
\end{equation}
where $T_0=2.725 \,\rm{K}$ is the CMB temperature at present time,
and $z_{LS}$ is the redshift at the surface of last scattering.
Despite of its small size, the ISW effect provides 
an independent test of dark energy, and as the effect is produced by 
a change in the gravitational potential, it could potentially be a new 
probe of modified gravity. We examine its impact on two observables: the 
CMB temperature anisotropy auto-correlation power spectrum, and the cross 
correlation between CMB anisotropy and the galaxy 
over-density along the line of sight.

\begin{figure}
\centering{
\epsfig{file=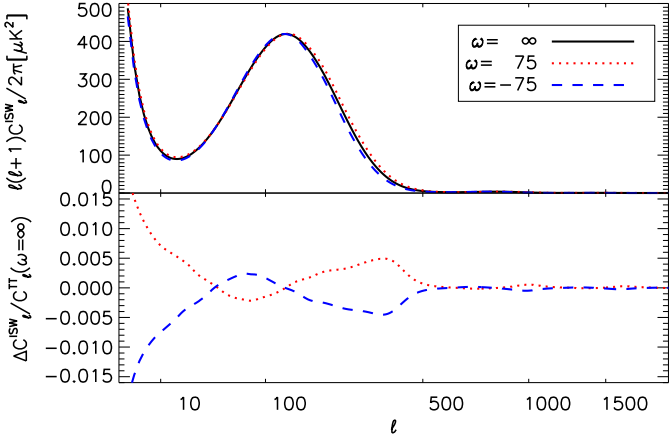,width=3.5in} \caption{The ISW
effect of TT power spectra in the scalar mode for the Brans-Dicke
gravity. $C_l^{ISW}$ is CMB TT power spectrum only considering the
ISW effect. $\Delta
C^{ISW}_l=C_l^{ISW}(\omega^\prime)-C_l^{ISW}(\omega=\infty)$,
$\omega^\prime=75$ for red dotted curve, and $\omega^\prime=-75$ for
blue dashed curve. } \label{fig:isw} }
\end{figure}

First we look at the CMB TT correlation. 
The $C_l^{ISW}$ spectra shown in Fig.\ref{fig:isw} are the temperature
anisotropy power spectra produced by the ISW effect, i.e. these are calculated
by including only the $\dot\Psi$ and $\dot\Phi$ (Newtonian gauge
variables) term in the line-of-sight integration with different models. 
According to the time when it occurred, the ISW effect could usually be
divided into two types: the early ISW effect during the 
radiation dominated to matter dominated transition, and the late
ISW effect during the matter dominated to dark energy dominated transition. 
The peaks of their contribution to 
the angular power spectrum have positions
corresponding to the respective horizon sizes. Thus, the early ISW effect 
produces the peak at $l \sim 150$, while the late ISW 
effect produces the slope at small $l$.

In the bottom panel of Fig.~\ref{fig:isw}, we plot $\Delta
C^{ISW}_l/C_l^{TT}(\omega=\infty)$, i.e. the ratio of the ISW
modification in the Brans-Dicke gravity to total TT power spectrum
of CMB in GR, where $\Delta
C^{ISW}_l\equiv C_l^{ISW}(\omega=75)-C_l^{ISW}(\omega=\infty)$ is plotted in 
red dotted curve, while $\omega=-75$ is plotted with 
the blue dashed curve. Here $C_l^{TT}(\omega=\infty)$ is the TT power spectrum
including all effects with $\omega=\infty$ (the GR case).
The correction of the late ISW effect (at the lowest $l$)  caused by the
Brans-Dicke theory is of the order of one percent of total TT power
spectrum, and the correction from the early ISW effect (at
$l\sim150$) is only about half of that size. 
This correction is buried in the cosmic
variance, and it would be hard to distinguish the Brans-Dicke gravity 
from General Relativity with this effect.

\begin{figure}
\epsfig{file=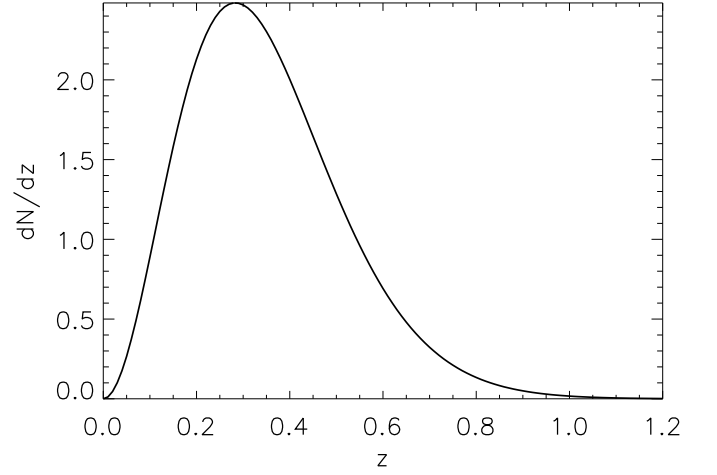,width=3.5in} \caption{ The redshift
distribution of the assumed sample with median redshift $z_m=0.33$
.} \label{fig:dndz}
\end{figure}

The cross correlation between CMB temperature and galaxy
over-density along the line of sight can also be used to measure the 
ISW effect. To calculate this effect, we consider
the observed galaxy density contrast in the direction
$\mathbf{\hat{n}}$,
 \bea \delta_{g}(\mathbf{\hat{n}}) = \int b_g(z) \frac{dN}{dz}(z)
 \delta_m(\mathbf{\hat{n}},z) dz \ .  \label{eq:delta_n}
 \eea
We assume that the bias is a constant, $b_g(z)=1.3$. The selection 
function $dN/dz$ describes the redshift
distribution of the galaxy sample,  here we adopt the analytic
function from Ref.\cite{cabre:2007}:
 \bea \frac{dN}{dz} \propto z^2 e^{-(z/z')^{3/2}} \ ,
 \eea
where $z'=z_m/1.412$, $z_m$ is the median redshift of the survey,  which we set
as $z_m=0.33$, and this galaxy redshift distribution is
shown in Fig.\ref{fig:dndz}, with the normalization of the distribution
satisfying $\int dN/dz=1$.
The gravitational potential $\Phi$ is related to the matter
density fluctuation $\delta$ via the Poisson equation:
\begin{equation}
\nabla^2 \Phi (\bmath{\hat{n}},z)= 4\pi G a^2 \rho_{\rm{m}}(z) \,
\delta (\bmath{\hat{n}},z) \ ,
\end{equation}
or
\begin{equation}
\Phi(\bmath{k},z)= -\frac{3}{2} \Omega_{\rm{m}} \left(\frac{H_0}{c
k}\right)^2 \frac{\delta(\bmath{k},z)}{a} \ , \label{equa:Phik}
\end{equation}
where $\rho_m=\rho_m^0 a^{-3}$.

The angular cross--correlation of the CMB temperature and galaxy
fluctuation is given by,
\bea w_{gT}^{\rm ISW}(\theta) &\equiv & {
\langle\delta_{g}(\bmath{\hat{n}_1})
\,\Delta_{T}(\bmath{\hat{n}_2})\rangle }  \\
&=& { \langle\delta_{g}(\bmath{\hat{n}_1}) \,\Delta_{T}^{ \rm
ISW}(\bmath{\hat{n}_2})\rangle }
 \eea
\noindent where $\bmath{\hat{n}_1} \bmath{\cdot}
\bmath{\hat{n}_2}=\cos{\theta}$. Note that
$\Delta_{T}(\bmath{\hat{n}})$ is the total temperature fluctuation at
given direction $\bmath{\hat{n}}$, while $\Delta_{T}^{\rm
ISW}(\bmath{\hat{n}})$ is the  temperature fluctuation caused only
by the ISW effect(seen Eq.(\ref{equa:isw})), the identities hold because
the CMB temperature fluctuations caused by other effects do not correlate 
with galaxy over-density. Expand
$\delta_m(\mathbf{\hat{n}},z)$ in Eq.(\ref{eq:delta_n}) into Fourier
modes:
 \bea
\delta_m(\mathbf{\hat{n}},z(\chi)) &=& \delta_m(\mathbf{\hat{n}}\chi,z)    \\
 &=& \int \frac{d^3 k}{(2\pi)^3}
\delta_m(k,z) e^{-i \rm{k} \cdot \mathbf{\hat{n}} \chi } \ ,
\label{equa:delta_m(K)}
 \eea
where $\chi$ is the comoving distance from redshift $0$ to $z$, and further
expand $e^{-i\bmath{k}\bmath{\cdot}\bmath{\hat{n}}\chi}$  as
\begin{equation}
e^{-i\bmath{k}\bmath{\cdot}\bmath{\hat{n}}\chi}=4\pi \sum_{lm}
(-i)^{l} \, j_l(k\chi) \, Y_{lm}(\bmath{\hat{n}}) \,
Y_{lm}^{\ast}(\bmath{\hat{k}}) \label{equa:plane} \ .
\end{equation}
where $j_l(x)$ is the spherical Bessel function of the first kind of
rank $l$, $Y_{lm}(\bmath{\hat{k}})$ is the spherical harmonic
function. Substituting Eqs.(\ref{equa:delta_m(K)}) and (\ref{equa:plane}) into
Eq.(\ref{eq:delta_n}), we obtain
 \bea
 \delta_{g}(\bmath{\hat{n}}) &=& \sum_{lm} \delta_{g,lm}
 Y_{lm}(\bmath{\hat{n}}) \ ,
\eea
where
\begin{eqnarray}
\delta_{g,lm}&=&(-i)^l \int \frac{d^3k}{(2\pi)^3} \int dz\,4\pi j_l(k\chi) \, Y^{\ast}_{lm}(\bmath{\hat{k}})  \nonumber \\
             & &   ~~~ \times  b_g(z) \, \frac{dN}{dz}(z) \,
             \delta(\bmath{k},z) \ .
\label{equa:glm}
\end{eqnarray}
Similarly, 
 \bea
  \Delta_{T}^{\rm{ISW}}(\bmath{\hat{n}}) &=&  \sum_{lm}  \Delta_{T, lm}^{\rm ISW} Y_{lm}(\bmath{\hat{n}}) \ ,
\eea
 where
\begin{eqnarray}
\Delta_{T,lm}^{\rm{ISW}}&=&(-i)^l \int \frac{d^3k}{(2\pi)^3} \int
dz\,4\pi j_l(k\chi(z)) \, Y^{\ast}_{lm}(\bmath{\hat{k}})  \nonumber
\\ & &   ~~\times 3 \Omega_{\rm{m}} T_{0}
\left(\frac{H_0}{kc}\right)^2 \frac{\partial}{\partial z}
\left[\frac{\delta(\bmath{k},z)}{a(z)}\right] \label{equa:Tlm} \ .
\end{eqnarray}
The angular cross correlation power spectrum of the galaxy
over-density and ISW temperature perturbation is then
\begin{equation}
C_{gT}^{\rm{ISW}}(l) \equiv  \delta_{l l'} \delta_{m m'} \langle
\delta_{g,lm} \, \Delta_{T,l'm'}^{\ast} \rangle. \label{equa:clfull}
\end{equation}
Using the small angle(large $l$,
$l\gg 1$) approximation for the spherical Bessel
functions\cite{Afshordi:2003xu}:
 \bea
 j_l(x)=\sqrt{\frac{\pi}{2 l +1}} \big[\delta_{\rm Dirac}(l+\frac{1}{2}-x) +\mathbf{O}(l^{-2})
 \big]  \ . \label{eq:j_l(x)}
 \eea
we have\cite{Ho:2008bz}
 \bea
 \frac{2}{\pi} \int k^2 dk j_l(k \chi) j_l(k
 \chi')=\frac{1}{\chi^2}\delta(\chi-\chi')) \ . \label{eq:int_j}
 \eea
With the linear growth factor  $D(z)$: $\delta(\bmath{k},z) = D(z) \,
\delta(\bmath{k},0)$ 
Under the Limber approximation\cite{Limber53,LoVerde:2008re}, 
we have
\begin{equation}
C_{gT}^{\rm{ISW}}(l) = \frac{4}{(2l+1)^2} \int dz \, P(k) \,
W_{\rm{ISW}}(z) \, W_g(z) \frac{H(z)}{c}.
\end{equation}
where $P(k)$ is the linear power spectrum at redshift zero,  $k
\approx {(l+1/2)/\chi(z)}$ obtained from Eq.(\ref{eq:j_l(x)}),
$W_{\rm{ISW}}(z)$ and $W_g(z)$ are the ISW and galaxy window
functions defined as
 \bea
 W_{\rm{ISW}}(z) &\equiv& 3 \Omega_m T_0 \left(\frac{H_0}{c}\right)^2
\frac{d}{dz} \left[\frac{D(z)}{a(z)}\right]
 \eea and
\begin{equation}
W_g(z) \equiv b_g(z) \, \frac{dN}{dz}(z) \, D(z)  \ .
\end{equation}
Finally, $w_{gT}^{\rm{ISW}}(\theta)$ is related to the cross--power
spectrum by the Legendre polynomials,
\begin{equation}
w^{\rm{ISW}}_{gT}(\theta) = \sum_{l=2}^{\infty} \frac{2l+1}{4\pi}\,
P_{l}(\cos\theta)\, C_{gT}^{\rm{ISW}}(l) \ , \label{equa:wgt}
\end{equation}
This summation does not include the monopole ($l=0$) and dipole ($l=1$) term,
as was done in the  \textit{WMAP} analysis \cite{Sawangwit:2009gd}.

\begin{figure}
\centering{ \epsfig{file=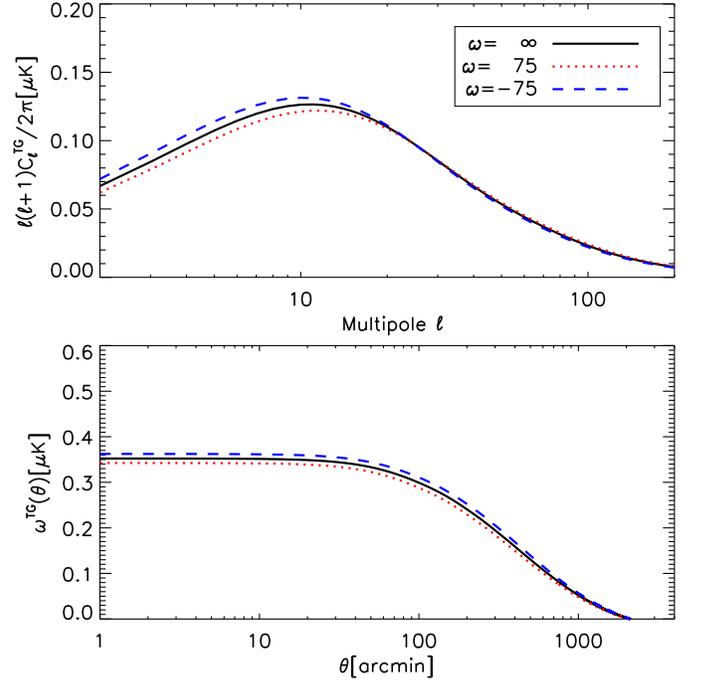,width=3.5in} \caption{The ISW
effect from CMB temperature and galaxy over-density correlation.
Upper panel is angular power spectrum, while bottom panel is angular
correlation function.} \label{fig:isw_cmb_g} }
\end{figure}

We plot the result in Fig.\ref{fig:isw_cmb_g}. The upper panel 
shows the angular power spectrum,  while the
bottom panel is the angular correlation function. 
For the angular power spectra, the Brans-Dicke theory with $\omega=\pm 75$
differs from the GR case by about 3-8 \% 
on large scales ($l<14$), and for the angular correlation
function, there is a difference of 3-4\% for $\theta<100$
arcmin, and on larger angles ($\theta>100$ arcmin) 
the difference is even larger. At present the 
CMB-galaxy correlation data could merely confirm the ISW
effect up to about 3 $\sigma$ level, and also often plagued by 
systematic errors which are not well-understood, as the 
observational results are often in conflict with each other 
\cite{Ho:2008bz,Sawangwit:2009gd,HernandezMonteagudo:2009fb,LopezCorredoira:2010rr}.
Recently, it is has been noted that for models in which the gravitational 
constant has drastic changes at low redshift, the ISW effect could be 
significant and thus provides a sensitive 
probe of the modified gravity 
\cite{Zhao:2010dz,Giannantonio:2009gi,Daniel:2010ky}. However, 
for the models discussed here, the variation of the gravitational potential 
at low redshift is actually not that large, thus 
including the ISW effect does not yield any significant difference.

\begin{figure}
\epsfig{file=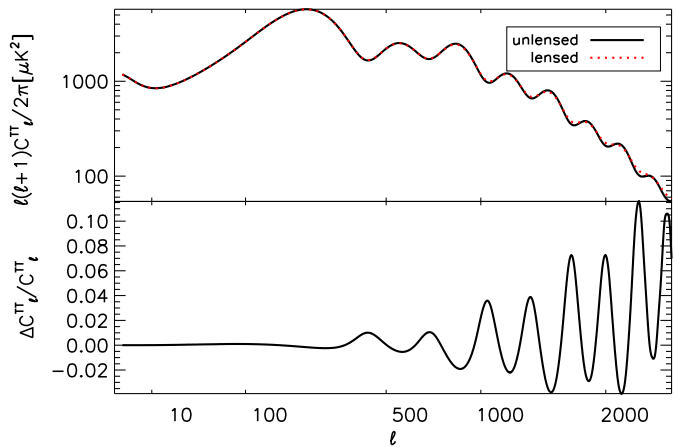,width=3.5in} \caption{Comparison of
unlensed and lensed CMB $TT$ power spectra in the scalar mode for
the Brans-Dicke gravity with $\omega=75$. $\Delta C^{TT}_l/C^{TT}_l$
is relative difference, where $\Delta C^{TT}_l= C^{TT
}_l(\mbox{lensed})- C^{TT }_l(\mbox{unlensed})$  and $C^{TT}$ in the
denominator is unlensed TT power spectrum.} \label{fig:lensing}
\end{figure}

How is the weak gravitational lensing effect modified in 
the Brans-Dicke theory? We investigate this problem by
modifying the CAMB code to include the lensing effect for the Brans-Dicke
theory.  The CMB lensing effect
in the Brans-Dicke gravity is similar to that in  the General
Relativity, it smooths the CMB power spectra in the small scales.
The result is showed in Fig.\ref{fig:lensing}, for $\omega=75$. 
At $l>500$, we begin to see corrections due to the Brans-Dicke theory 
at the percent level, so in the future when the Planck data becomes available, 
this effect should be included in the calculation. At 
$l>2000$, the correction could reach as high as ten
percent level, but there the CMB primordial anisotropy
is strongly damped, and the anisotropy is dominated by the SZ effect.

\section{Conclusion and summary}

Compared with Einstein's general relativity, there is an additional
scalar field coupled with the Ricci scalar in the Brans-Dicke
gravity, which makes the perturbation theory more complicated.
 With a covariant 1+3 approach, we have developed  a full set of
covariant and gauge-invariant formalism for calculating the cosmic
microwave background temperature and polarization anisotropies in
the Brans-Dicke gravity.  Instead of using the components of metric
as basic variables, the covariant formalism performs a 1+3 split of
the Bianchi and Ricci identities, using the kinematic quantities,
energy-momentum tensors of the fluid(s) and the gravito-
electromagnetic parts of the Weyl tensor to study how perturbations
evolve.  Adopting covariantly defined, gauge-invariant variables
throughout ensures that in our discussion the gauge ambiguities
is avoided, and all variables had a clear,
physical interpretation.   Since the definition of the covariant
variables does not assume any linearization, exact equations can be
found for their evolution, which can then be linearized around the
chosen background model. Furthermore, unified treatment of scalar,
vector and tensor modes do not require decomposing the different
modes from beginning as done in the metric method.
A price we have to pay is that with this method the calculation is
more complicated.

We then calculate the CMB temperature and polarization spectra  for
the Brans-Dicke models using a modified CAMB code.  In this paper we
consider both the scalar modes and the tensor modes in  adiabatic initial
condition, and adopt $\varphi_0=(2\omega+4)/(2\omega+3)$ at the
current epoch and $\varphi'=0$ at early time as initial condition
of the Brans-Dicke field.  Compared with the general-relativistic
model with the same cosmological parameters, both the amplitude and
the width of the acoustic peaks are different in the Brans-Dicke
models. We find that the small scale spectra and the polarization
spectra will provide a sensitive and vigorous constraint on the
different Brans-Dicke models in the scalar mode. For tensor modes, the
largest difference in CMB spectra for various Brans-Dicke models
are located at  $l\sim80$. The structure
formation process in the Brans-Dicke theory is also studied. The
matter power spectrum is shown in Fig.\ref{fig:spectrum.matter.z0}.
For positive $\omega$ case, the bend of the matter power spectra
occurs at shorter wavelengths, and there is thus more small-scale
power compared with the General Relativity case.

The ISW effect of the Brans-Dicke theory is investigated (see
Fig.\ref{fig:isw}) and Fig.\ref{fig:isw_cmb_g}). 
The correction to total TT power spectra come
from the early ISW effect caused by the Brans-Dicke theory is proved
to be of the order of one percent, and the late ISW effect is only
a half of the early ISW effect.  Due to the large cosmic variance at the
large scales, this effect is not significant in observational constraint. 
For CMB-galaxy
cross-correlation, the differences between the GR case and the Brans-Dicke case
with $\omega=\pm 75$ are at the 3-8 \% level
on large scales ($l<14$) in angular power spectra, or
3-4\% in angular correlation function for $\theta<100$ arcmin, and 
on even larger angular scales ($\theta>100$ arcmin) the difference is still 
larger. Nevertheless, for the modified gravity 
model considered here, where the variation in gravitational potential at 
low redshift is not very large, the ISW effect does not
provide a very sensitive probe due to the large cosmic covariances.

The CMB lensing
effect is plotted in Fig.\ref{fig:lensing}. This effect only appears
significantly at $l>2000$. The lensed CMB power spectra look smooth
at the small scale compared with the unlensed power spectra in the
Brans-Dicke gravity, which is very similar with the case in General
Relativity.

Our covariant calculation for the Brans-Dicke model 
is generally in agreement with previous results obtained in particular
gauges (e.g. the synchronous gauge\cite{chenxuelei:1999brans}). Furthermore, 
we have also obtained for the first time the temperature and polarization
spectra for tensor mode perturbations, the ISW effect, 
and the CMB lensing effect in Brans-Dicke theory.
The structure and speed of the code 
are greatly improved, thus providing a more powerful and convenient
tool for further studies. In paper II, we use the code and MCMC
algorithm to derive the constraint on the Brans-Dicke parameter
$\omega$ with the latest CMB and LSS observational data.

As the final remark, the covariant approach and corresponding CMB
code for the Brans-Dicke theory developed in this paper, together
with the synchronous gauge approach and corresponding code developed
in the previous paper\cite{chenxuelei:1999brans}, provide
consistent, systematic and complete methods to study the Brans-Dicke
theory. These methods and codes could be generalized to study more
general scalar-tensor theory, as well as more complex initial
condition, we plan to carry out such generalization in subsequent
studies.

\appendix

\section*{Acknowledgements}

We thank  Antony Lewis, G.F.R. Ellis, Marc Kamionkowski and Gong-Bo Zhao
for helpful discussions.
Our MCMC chain computation was performed on the Supercomputing Center of
the Chinese Academy of Sciences and the Shanghai Supercomputing
Center. X.C. acknowledges the hospitality of the Moore center of theoretical
cosmology and physics at Caltech, where part of this research is performed.
This work is supported by
the National Science Foundation of China under the Distinguished Young
Scholar Grant 10525314, the Key Project Grant 10533010; by the
Chinese Academy of Sciences under grant KJCX3-SYW-N2; and by the
Ministry of Science and Technology under the National Basic Science
program (project 973) grant 2007CB815401.

%\section*{References}

\bibliography{brans}
\bibliographystyle{apsrev}

\end{document}